\documentclass[12pt]{elsarticle}
\usepackage{amsmath,amssymb,bm}
\usepackage{paralist}
\usepackage{graphicx} 
\usepackage{epsfig} 
\usepackage{subfigure,appendix}
\usepackage{psfrag}
\usepackage{multicol}
\usepackage{setspace,tabularx,fancyhdr}
\usepackage{ulem,natbib}
\usepackage{color}
\usepackage[usenames,dvipsnames,svgnames,table]{xcolor}
\usepackage[top=1in,bottom=1in,left=1in,right=1in]{geometry}
\usepackage{framed}
\usepackage{xcolor}
\usepackage{placeins}
\usepackage{nicefrac} 
\usepackage{url} 
\usepackage{subfloat}

\title{
On the two-dimensional extension of one-dimensional algebraically growing waves at neutral stability
}
\author[Label1]{Colin M. Huber}
\author[Label1]{Nathaniel S. Barlow}
\author[Label1,Label2]{Steven J. Weinstein}

\address[Label1]{School of Mathematical Sciences,
			Rochester Institute of Technology, 
           	Rochester,
            NY,
            14623, 
            USA}
\address[Label2]{Department of Chemical Engineering,
            Rochester Institute of Technology, 
            Rochester, 
            NY,
			14623,            
            USA}

\date{Spring, 2022}
\begin{document}
\begin{frontmatter}

\begin{abstract}
This work considers two linear operators which yield wave modes that are classified as neutrally stable, yet have responses that grow or decay in time. Previously, King et al. (Phys. Rev. Fluids, 1, 2016, 073604:1-19) and Huber et al. (IMA J. Appl. Math., 85, 2020, 309-340) examined the one-dimensional (1D) wave propagation governed by these operators. Here, we extend the linear operators to two spatial dimensions (2D) and examine the resulting solutions. We find that the increase of dimension leads to long-time behaviour where the magnitude is  reduced by a factor of $t^{\nicefrac{-1}{2}}$ from the 1D solutions. Thus, regions of the solution which grew algebraically as $t^{\nicefrac{1}{2}}$ in 1D now are algebraically neutral in 2D, whereas regions which decay (algebraically or exponentially) in 1D now decay more quickly in 2D. Additionally, we find that these two linear operators admit long-time solutions that are functions of the same similarity variable that contracts space and time.
\end{abstract}

\end{frontmatter}


\section{Introduction}
	\label{sec:General:Introduction}

An essential feature of a partial differential equation operator that models wave propagation is whether it admits solutions that will amplify or damp with time as waves propagate in space. In some physical processes, such as in liquid fuel atomizers~\cite{ibrahim1995,lin2003,el2015}, instabilities are necessary to enable the flow to break into droplets, and in other cases turbulence is used to facilitate mixing~\cite{paul2004}.  In many processes, however, stability of wave-like disturbances is not only desired, it is required, such as in thin film coating used to manufacture printed electronics and liquid crystal display screens~\cite{cohen1992,kistler1997}.
Although the latter processes are governed by the nonlinear equations of fluid dynamics, it is standard to linearize about steady operating conditions; coating processes are run with tight constraints on the uniformity of the product, and thus linearized equations are often sufficient to predict the effect of process disturbances~\cite{weinstein2004}. 

Classical stability theory, introduced by Lord Rayleigh~\cite{rayleigh1880} in 1880 and further refined over the next hundred years, serves to classify the response, $h$, of a linearized system by assembling its fundamental modes (responses), $h_\textbf{k}$, as $h = \sum h_\textbf{k}$. For a two-dimensional (2D) linear operator, each mode may be expressed as:

\begin{equation}
\label{eq:general:ModalForm}
h_\textbf{k}
=  A(\textbf{k}) e^{i(\textbf{k}\cdot\textbf{x}-\omega t)} 
= A(k_x,k_z) e^{i(k_xx+k_zz-\omega t)}
\end{equation}
where $\textbf{k}$ is the real wave number vector, $\textbf{x}$ is the spatial vector, $\omega$ is a complex frequency, $t$ is time, and $A(\textbf{k})=A(k_x,k_z)$ is the wave number dependent amplitude. 
By substituting Eq.~(\ref{eq:general:ModalForm}) into the homogeneous version of the linearized operator, the dispersion relation $\omega\equiv\omega(\textbf{k})=\omega(k_x,k_z)$ is found such that the result --Eq.~(\ref{eq:general:ModalForm})-- is nontrivial (i.e., leaving $A(\textbf{k})$ arbitrary). The idea, then, is to examine the behaviour of individual modes to draw stability conclusions. In particular,
since $\textbf{k}$ is real valued, the only mechanism for Eq.~(\ref{eq:general:ModalForm}) to admit exponential growth or decay is through the imaginary part of the complex frequency, $\omega_i(\textbf{k})\equiv\text{Im}[\omega(\textbf{k})]$.
Any single mode that grows in time will dominate other modes which decay in time. 
Furthermore, of the growing modes, the one that grows the fastest (or decays the slowest) will dominate the behaviour of $h$ as $t\to\infty$ with an exponential growth (or decay) rate of $\omega_{i,\mathrm{max}}\equiv\text{max}(\omega_i(\textbf{k}))$ determined from the dispersion relation.
The amplitude of the long-time behaviour, $h_\mathrm{max}$, may be expressed as:

\begin{equation}
\label{eq:general:hMax}
h_\mathrm{max}\sim 
A_\mathrm{max} e^{\omega_{i,\mathrm{max}}t}
\qquad
\text{ as } t\to\infty,
\end{equation}
where $A_\mathrm{max}$ is the amplitude of the maximum growth mode. In Eq.~(\ref{eq:general:hMax}), $\omega_{i,\mathrm{max}}$ can be used to determine the classical stability for a given set of conditions as follows~\cite{Chandrasekhar, HuerreRossi, huerre2000}:

\begin{subequations}
\label{eq:general:ClassicalStabilityClassification}

\begin{align}
\label{eq:general:Stability}
\omega_{i,max}<0:&\quad\text{ The system is stable}\\  
\label{eq:general:NeutralStability}
\omega_{i,max}=0:&\quad\text{ The system is neutrally stable}\\
\label{eq:general:Instability}
\omega_{i,max}>0:&\quad\text{ The system is unstable}
\end{align}
\end{subequations}
Depending on the operator being studied, the stability of the system may depend on the parameters in the governing equation. If a given set of parameters results in $\omega_{i,\mathrm{max}}=0$, the condition of neutral stability is met. 
When this occurs, the system governed by the operator is said to be at the ``neutral stability boundary", as small changes in the parameters can often move the system to a state of stability  or instability .

Classical stability analysis implies that a stable system --Eq.~(\ref{eq:general:Stability})-- will exhibit exponentially damped responses with time whereas a system that is unstable --Eq.~(\ref{eq:general:Instability})-- will exhibit exponential amplification. However, at the boundary between the two regimes --Eq.~(\ref{eq:general:NeutralStability})--, the method fails to accurately predict the long-time linear stability of the system, as shown in the previous work by King et al.~\cite{king2016} and Huber et al.~\cite{Huber2020}. Specifically, King et al. examined a one-dimensional (1D) operator (henceforth referred to as 1D-KRK)\footnote{Notation is chosen to reflect the dimesionality of the operator and the lead author's initials} that governs the response to varicose perturbations in a curtain flow, and Huber et al. studied a 1D operator (henceforth referred to as 1D-CMH)\footnotemark[\value{footnote}] that enabled the neutral stability threshold in Eq.~(\ref{eq:general:ClassicalStabilityClassification}) to be traversed by the variation of a parameter. At the neutral stability boundary, neither operator should result in growing or decaying responses as per the classification given by Eq.~(\ref{eq:general:NeutralStability}); however, both operators admit solutions that exhibit \textit{algebraic} (i.e., $t^p$, $p$ real and rational) growth in time. In this paper, we examine the 2D extensions of the 1D-KRK and 1D-CMH operators, denoted respectively as the 2D-KRK and 2D-CMH operators.

A closely related set of problems is that of one-dimensional water waves that are neutrally stable according to Eq.~(\ref{eq:general:NeutralStability}) but whose amplitudes damp with an algebraic $t^{-\frac{1}{2}}$ dependence at long times~\cite{whitham2011, lighthill2001}. Lighthill notes that the damping rate is reduced by a factor of $t^{-\frac{1}{2}}$ with the addition of each spatial dimension~\cite{lighthill2001}. 
Herein, we show that this feature is also true of at least two systems that exhibit algebraic growth in 1D.

The paper is organized as follows. In Section~\ref{sec:KP:KRK}, the 2D-KRK operator is introduced, its Fourier integral solution is obtained, and its long-time asymptotic behaviour is determined. Section~\ref{sec:NP:CMH} similarly examines the 2D-CMH operator. A comparison between the 2D responses in Sections~\ref{sec:KP:KRK} and~\ref{sec:NP:CMH} and their 1D counterparts in~\cite{king2016} and~\cite{Huber2020} is provided in Section~\ref{sec:General:Discussion}, and concluding remarks are provided in Section~\ref{sec:General:Conclusions}.

\section{2D-KRK: 2D extension of algebraically growing 1D-KRK model for varicose waves in liquid curtains}
\label{sec:KP:KRK}

\subsection{Problem statement}

The first problem examined here is the 2D extension of a 1D operator derived from a model of varicose waves in a thin flowing curtain in the absence gravity and with passive ambient gas~\cite{lin2003}. The previous 1D analysis found that the response to disturbances in this flow grows algebraically; a natural extension is to examine the response to a 2D disturbance. The increase in dimension is implemented by replacing all the 1D spatial derivatives with 2D del operators as follows:

\begin{subequations}
\label{eq:KP:Operator}

\begin{equation}
\label{eq:KP:LiteralOperator}
\left[
\left(
\frac{\partial}{\partial t}+
\textbf{c}\cdot\nabla
\right)^2
+B^2\nabla^4
\right]h=
A\delta(x)\delta(z)\delta(t),
\end{equation}

%

\begin{equation}
\label{eq:KP:DelOperators}
\textbf{c}\cdot\nabla =
c_x\frac{\partial}{\partial x}+
c_z\frac{\partial}{\partial z},
\qquad
\nabla^4 =
\frac{\partial^4}{\partial x^4}+
2\frac{\partial^4}{\partial x^2z^2}+
\frac{\partial^4}{\partial z^4}
\end{equation}

\begin{equation}
\label{eq:KP:InitialConditions}
h(x,z,0)=0
,\qquad
\frac{\partial h}{\partial t}(x,z,0)=0
\end{equation}

\begin{equation}
\label{eq:KP:OtherConditions}
h\to 0
\quad\text{as}\quad
x,z\to\pm\infty
,\qquad
c_x,c_z \geq 0
,\qquad
B,A \in \mathbb{R}.
\end{equation}  
\end{subequations}
In Eq.~(\ref{eq:KP:LiteralOperator}), $\textbf{c}$ is the underlying convective fluid flow vector with components $c_x$ and $c_z$ and $B$ is a real valued parameter that is related to the Weber number as described in~\cite{king2016}. Note that the forcing function in Eq.~(\ref{eq:KP:LiteralOperator}) has the real-valued amplitude $A$. As written, the constraints in Eq.~(\ref{eq:KP:InitialConditions}) are chosen to be homogeneous; care, however, has been taken in making that choice. It is clear from the 1D-KRK and the 1D-CMH \cite{king2016,Huber2020} analyses that the form of initial conditions can affect whether algebraic growth occurs. This is discussed further in the next section, before proceeding to examine the solution to the system in Eq.~(\ref{eq:KP:Operator}).

\subsection{Initial Condition Justification}

It is first useful to note that an impulse disturbance to the surface velocity, $\nicefrac{\partial h}{\partial t}$ in Eq.~(\ref{eq:KP:InitialConditions}), has the exact same effect as the impulse forcing function included in Eq.~(\ref{eq:KP:LiteralOperator}) (see~\ref{app:KP:InitialConditions} for details). For this reason, the initial velocity may be taken to be zero in Eq.~(\ref{eq:KP:InitialConditions}) without loss of generality.
Additionally, an impulse disturbance to initial surface height, $h$ in Eq.~(\ref{eq:KP:InitialConditions}), leads to a solution that violates the condition of $h\to 0$ as $x,z\to\pm\infty$ in Eq.~(\ref{eq:KP:OtherConditions}); this nonphysical solution is included in~\ref{app:KP:InitialConditions}. Although not stated explicitly in~\cite{king2016}, this violation also occurs in 1D-KRK.
An interesting feature of this 2D nonphysical solution is that, because the Fourier transform of the solution exists through rapid oscillations of the sinusoidal integrand, the Fourier integral solution methodology admits the violation.
If one extends the class of allowable solutions to include spatially non-local responses, the solution nevertheless does damp with time. To avoid the nonphysical nature of this solution however, the initial height is chosen to be zero in Eq.~(\ref{eq:KP:InitialConditions}). In summary, homogeneous constraints on the system in Eq.~(\ref{eq:KP:Operator}) are chosen without loss of generality in the stability conclusions that follow.

\subsection{Classical stability analysis}

In the one-dimensional problem 1D-KRK, all modes $h_\textbf{k}$ in Eq.~(\ref{eq:general:ModalForm}) are neutrally stable~\cite{king2016}. 
The addition of a higher dimension does not change this behaviour. Through the substitution of the modal form --Eq.~(\ref{eq:general:ModalForm})-- into the homogeneous version of Eq.~(\ref{eq:KP:LiteralOperator}), the following dispersion relation is obtained:

\begin{equation}
\label{eq:KP:DispersionRelation}
\omega = (c_x k_x+c_z k_z) \pm B (k_x^2+k_z^2).
\end{equation}
In Eq.~(\ref{eq:KP:DispersionRelation}), $k_x$ and $k_z$ are the real wave numbers in the $x$ and $z$ directions, respectively, and $\omega$ is the complex frequency.
Because $c_x, k_x,c_z,k_z$ are real, $\omega$ is always real, just as in the case of water waves~\cite{whitham2011, lighthill2001}. As such, the imaginary part of $\omega$ is always $0$, and all modes --Eq~(\ref{eq:general:ModalForm})-- are neutrally stable for all values of the parameters according to Eq.~(\ref{eq:general:ClassicalStabilityClassification}).
As a check, note that the 1D dispersion relation (Equation (5a) in~\cite{king2016} with $n=2$) is recovered by letting $c_z=k_z=0$ in Eq.~(\ref{eq:KP:DispersionRelation}).
According to classical stability analysis  in Eq.~(\ref{eq:general:ClassicalStabilityClassification}), the response to such a disturbance should neither grow nor decay. However, the solution grows algebraically in 1D as shown in King et al.~\cite{king2016}.

\subsection{Integral solution}
\label{sec:KP:IntegralSolution}
The solution to Eq.~(\ref{eq:KP:Operator})  is found by taking Fourier transforms in $x$ and $z$ (resulting in the transformed variable $\hat{\hat{h}}_{xz}$) and the Laplace transform in $t$. The resulting Fourier inversion integral solution is given as

\begin{equation*}
h(x,z,t) = \frac{1}{4\pi^2}
	\int\limits_{-\infty}^{\infty} 
	\int\limits_{-\infty}^{\infty} 
	\hat{\hat{h}}_{xz}
	e^{ik_xx}
	e^{ik_zz}
	dk_xdk_z ,
\end{equation*}

\begin{equation}\label{eq:KP:FourierSolution}
\hat{\hat{h}}_{xz}=
Ae^{-i\psi t}
\frac{\sin\left(\eta t\right)}{\eta},
\qquad
\psi = c_xk_x + c_zk_z,
\qquad
\eta = B(k_x^2+k_z^2). 
\end{equation}
Note that there is a removable singularity at $\eta=0$ in Eq.~(\ref{eq:KP:FourierSolution}). In order to extract the asymptotic behaviour of the integral at large times, the standard approach is to express the sine function in terms of complex exponentials and evaluate each new integral as $t$ goes to infinity. However, doing so leads to poles along the path of integration, i.e. principal values. Although one could proceed in this way, the issue is avoided entirely by introducing the new variable $\xi$ into the integral --Eq.~(\ref{eq:KP:FourierSolution})-- to obtain

\begin{subequations}
\begin{equation}
\label{eq:KP:XiIntegral}
\tilde{h}(x,z,t,\xi) = 
\frac{A}{4\pi^2}
\int\limits_{-\infty}^{\infty}
\int\limits_{-\infty}^{\infty}
e^{-i\psi t}\left(
\frac{\sin(\xi \eta t)}
{\eta}
\right)
e^{ik_xx}
e^{ik_zz}
dk_zdk_x,
\end{equation}

\begin{equation}
\label{eq:KP:Relations}
\tilde{h}(x,z,t,0) = 0,
\qquad
\tilde{h}(x,z,t,1) = h(x,z,t),
\end{equation}
\end{subequations}
where Eqs.~(\ref{eq:KP:FourierSolution}) and~(\ref{eq:KP:XiIntegral}) are linked via the relation in Eq.~(\ref{eq:KP:Relations}).
The structure of the integrand in Eq.~(\ref{eq:KP:XiIntegral}) is chosen such that taking its derivative with respect to $\xi$ completely removes the $\eta$ in the denominator as:

\begin{equation}
\label{eq:KP:DerivativeIntegral}
\frac{\partial\tilde{h}}{\partial \xi} = 
\frac{At}{4\pi^2}
\int\limits_{-\infty}^{\infty}
\int\limits_{-\infty}^{\infty}
e^{-i\psi t}\cos(\xi \eta t)
e^{ik_xx}
e^{ik_zz}
dk_zdk_x.
\end{equation}
The solution for $h$ may then be obtained via the integration of the resulting Eq.~(\ref{eq:KP:DerivativeIntegral}) using using the constraints in Eq.~(\ref{eq:KP:Relations}):

\begin{equation}
\label{eq:KP:IntegralInXi}
h(x,z,t) = \int\limits_0^1 \frac{\partial\tilde{h}}{\partial\xi}d\xi.
\end{equation}	
The cosine in Eq.~(\ref{eq:KP:DerivativeIntegral}) is complexified and the resulting integral is decomposed as

\begin{subequations}
\label{eq:KP:QWSplit}
\begin{equation}
\label{eq:KP:DerivativeQWIntegral}
\frac{\partial\tilde{h}}{\partial \xi} = \frac{At}{8\pi^2}
\left(
\mathcal{Q}+\mathcal{W}
\right),
\end{equation}

\begin{equation}
\label{eq:KP:QIntegral}
\mathcal{Q}=
\int\limits_{-\infty}^{\infty}
e^{\left(i\xi Bk_z^2+i(\frac{z}{t}-c_z)k_z \right)t}
dk_z
\left[
\int\limits_{-\infty}^{\infty}
e^{\left(i\xi Bk_x^2+i(\frac{x}{t}-c_x)k_x \right)t}
dk_x
\right],
\end{equation}

\begin{equation}
\label{eq:KP:WIntegral}
\mathcal{W}=
\int\limits_{-\infty}^{\infty}
e^{\left(-i\xi Bk_z^2+i(\frac{z}{t}-c_z)k_z\right)t}
dk_z
\left[
\int\limits_{-\infty}^{\infty}
e^{\left(-i\xi Bk_x^2+i(\frac{x}{t}-c_x)k_x \right)t}
dk_x
\right].
\end{equation}
\end{subequations}
The integrals in Eqs.~(\ref{eq:KP:QIntegral}) and~(\ref{eq:KP:WIntegral}) are evaluated in closed form (see~\ref{app:KP:ContourIntegration}) to yield:

\begin{equation}
\label{eq:KP:Derivative}
\frac{\partial\tilde{h}}{\partial \xi} =
\frac{A }{4\xi B\pi }
\sin\left(
\frac{t}{4\xi B}\left(
(V_x-c_x)^2
+(V_z-c_z)^2
\right)
\right)
\end{equation}
where $V_x = \nicefrac{x}{t}$ and $V_z=\nicefrac{z}{t}$ are velocities introduced to provide a convenient representation of the solution.

Eq.~(\ref{eq:KP:Derivative}) is then substituted into Eq.~(\ref{eq:KP:IntegralInXi}) and integrated to attain the solution to Eq.~(\ref{eq:KP:Operator}). The exact solution can be expressed in terms of a similarity variable $\Phi$ as

\begin{subequations}
\label{eq:KP:WholeSolution}
\begin{equation}
\label{eq:KP:Solution}
h(\hat{V},t)=
\frac{A}{4B\pi}
\left(
\frac{\pi}{2}-
\int\limits_{0}^{\frac{\Phi}{4B}}
\frac{1}{u}
\sin(u)du
\right),
\qquad
\Phi = \hat{V}^2t,
\end{equation}

\begin{equation}
\label{eq:KP:VHatDef}
\hat{V}=\sqrt{(V_x-c_x)^2+(V_z-c_z)^2}.
\end{equation}
\end{subequations}
\noindent
In Eq.~(\ref{eq:KP:VHatDef}), $\hat{V}=0$ corresponds to the peak of the response. More generally, $\hat{V}$ is defined as the locus of velocities relative to the velocity of the convecting peak $(c_x,c_z)$. Note that when $\hat{V}=0$ in Eq.~(\ref{eq:KP:Solution}) the solution has a constant height expressed as

\begin{equation}
\label{eq:KP:PeakSolution}
h\big|_{\hat{V}=0}=
\frac{A}{8B}.
\end{equation}
The long-time asymptotic behaviour of the sine integral in Eq.~(\ref{eq:KP:Solution}) for $\hat{V}\neq 0$ may be expressed as:

\begin{equation}
\label{eq:KP:Asymptotic}
h(\Phi)\Big|_{(\hat{V}\neq 0)} \sim
\frac{A}{\pi\Phi}
	\cos\left(
		\frac{\Phi}{4B}
	\right)
+O\left(\frac{1}{\Phi^2}\right),
\quad \Phi=\hat{V}^2t,
\quad\text{ as }
\Phi\to\infty.
\end{equation}
The limit of $\Phi$ going to infinity aligns with the limit of $t$ going to infinity for any fixed non-zero $\hat{V}$.

\begin{figure}[ht!]
\centering
\includegraphics[keepaspectratio,width=7in]{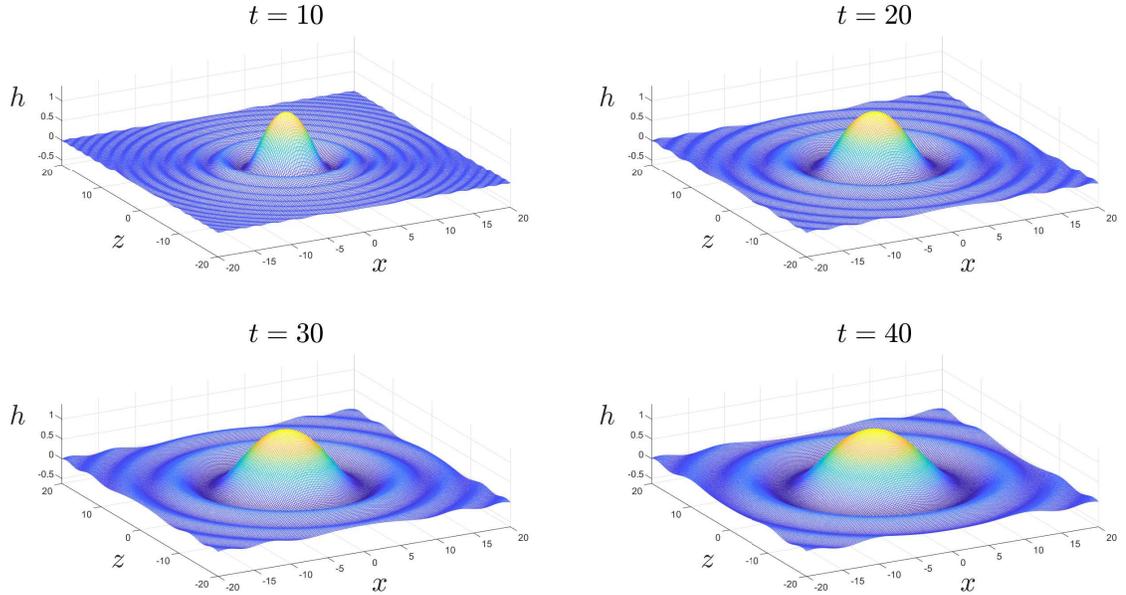}
\caption{Time evolution of 2D-KRK response given by Eq.~(\ref{eq:KP:Solution}) with $A=10$ and $B=1$. There is no underlying convective flow as $c_x=c_z=0$. The constant height peak remains situated at $x=z=0$ and the breadth of the response increases as time progresses. The radius of any given phase feature (e.g. crest or trough) increases as $\sqrt{t}$ as given by Eq.~(\ref{eq:KP:RingsOfConstantHeightNonconvective}).
}
\label{fig:KP:3D_no_convection}
\end{figure}

\begin{figure}[ht!]
\centering
\includegraphics[keepaspectratio,width=7in]{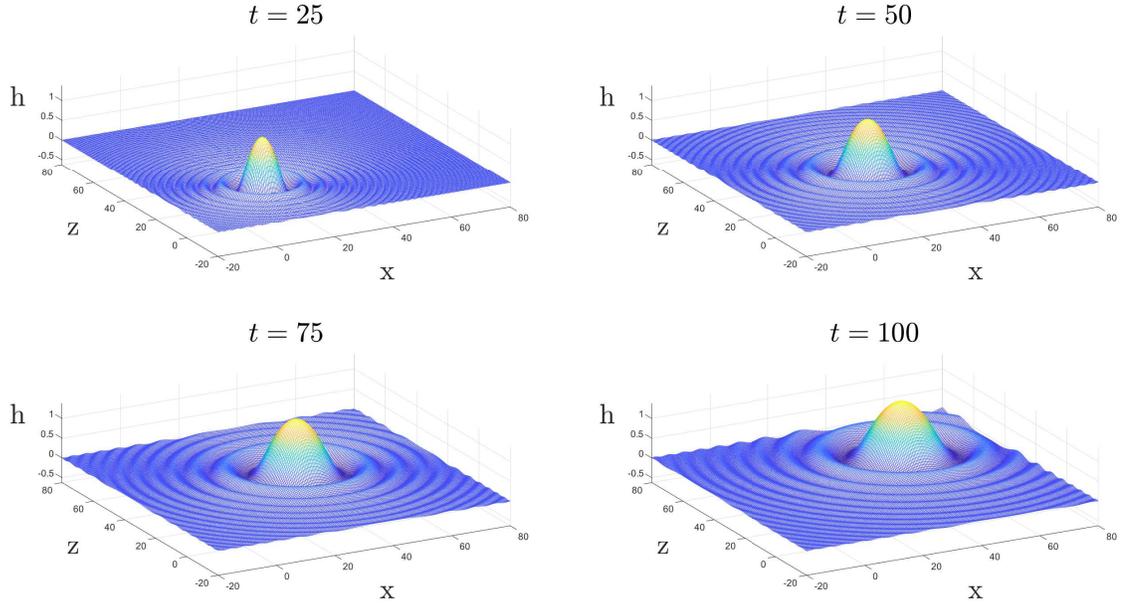}
\caption{Time evolution of 2D-KRK response given by Eq.~(\ref{eq:KP:Solution}) with $A=10$, $B=1$, and underlying convective flow of $c_x=c_z=0.5$. The peak moves along $(x,z)=(c_xt,c_zt)$ , and the breadth of any given phase feature (e.g. crest or trough) propagates radially relative to this position with time as $\sqrt{t}$ as given by Eq.~(\ref{eq:KP:RingsOfConstantHeight}).
}
\label{fig:KP:3D_convection}
\end{figure}

The sine integral solution --Eq.~(\ref{eq:KP:Solution})-- (implemented using MATLAB's sinint function) is shown in Figs.~\ref{fig:KP:3D_no_convection} and~\ref{fig:KP:3D_convection} for different values of the convection parameters $c_x$ and $c_z$. In accordance with Eq.~(\ref{eq:KP:PeakSolution}), the height of the response peak (i.e. $\hat{V}=0$) remains constant as the response evolves in time. For the other loci of velocities, $\hat{V}\neq 0$, the solution decays as $O(\nicefrac{1}{t})$ in accordance with Eq.~(\ref{eq:KP:Asymptotic}).

\FloatBarrier

\begin{figure}[ht!]
\centering
\includegraphics[keepaspectratio,width=6in]{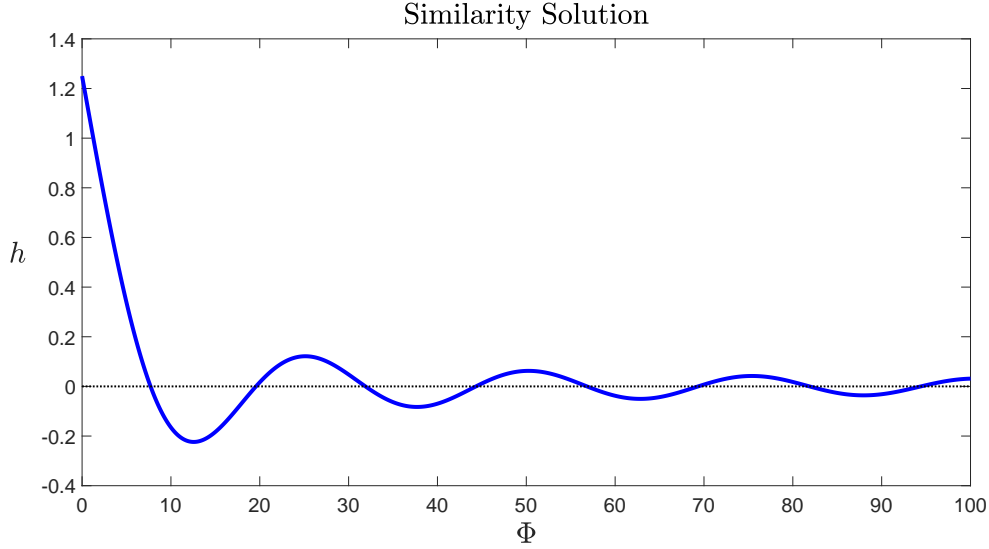}
\caption{Similarity solution --Eq.~(\ref{eq:KP:Solution})-- for $A=10$ and $B=1$. 
}
\label{fig:KP:SimilaritySolutionOnly}
\end{figure}

\FloatBarrier
\noindent
The similarity variable in Eq.~(\ref{eq:KP:Solution}) implies that a similarity transform could be applied to Eq.~(\ref{eq:KP:Operator}) to attain an ordinary differential equation whose exact solution is Eq.~(\ref{eq:KP:Solution}).
Fig.~\ref{fig:KP:SimilaritySolutionOnly} provides a plot of the similarity solution, which has been represented spatially in Figs.~\ref{fig:KP:3D_no_convection} and~\ref{fig:KP:3D_convection}.  By inspection, it is observed that the spatial responses of Figs.~\ref{fig:KP:3D_no_convection} and~\ref{fig:KP:3D_convection} are rotations and stretches of this solution relative to the location of the traveling peak ($\hat{V}=0$) at $x=c_xt$ and $z=c_zt$ according to Eq.~(\ref{eq:KP:VHatDef}).

\FloatBarrier

The similarity solution --Eq.~(\ref{eq:KP:Solution})-- may be used to extract more features of the solution in the physical ($x,z,t$) domain.
If $\hat{V}$ is held fixed in Eq.~(\ref{eq:KP:WholeSolution}), then $\Phi$ goes as $t$, and we can examine the behaviour of any propagation speed relative to that of the peak (corresponding to $\hat{V}=0$).  According to Eq.~(\ref{eq:KP:Asymptotic}), then, the response for any fixed $\hat{V}$ decays asymptotically as $\nicefrac{1}{t}$ for large times.
The variable $\Phi$ may be equivalently expressed explicitly in terms of space and time as:

\begin{equation}
\label{eq:KP:PhiDef}
\Phi = \frac{R^2}{t},
\quad
R^2=(x-c_xt)^2+(z-c_zt)^2,
\end{equation}
where $R$ is the radius of a circle centered at $x=c_xt$ and $z=c_zt$. Thus for fixed $R$, $\Phi$ will go as $\nicefrac{1}{t}$ as shown in Eq.~(\ref{eq:KP:PhiDef}). A fixed value of $R$ provides the locus of points a distance of $R$ away from the convecting peak. This is most easily visualized when $c_x=c_z=0$ (Fig.~\ref{fig:KP:3D_no_convection}), as fixed values of $R$ corresponds to fixed concentric circles in the $x,z$ plane. As time goes on, the solution at a fixed $R$ will yield smaller $\Phi$ values; Fig.~\ref{fig:KP:SimilaritySolutionOnly} indicates that the solution grows towards the maximum value of the peak as $t\to\infty$ (i.e. $\Phi\to 0$ for fixed $R$).

\subsection{Spatial stability}
	\label{sec:KP:SpatialStability}
\begin{figure}[ht!]
\centering
\includegraphics[keepaspectratio,width=6in]{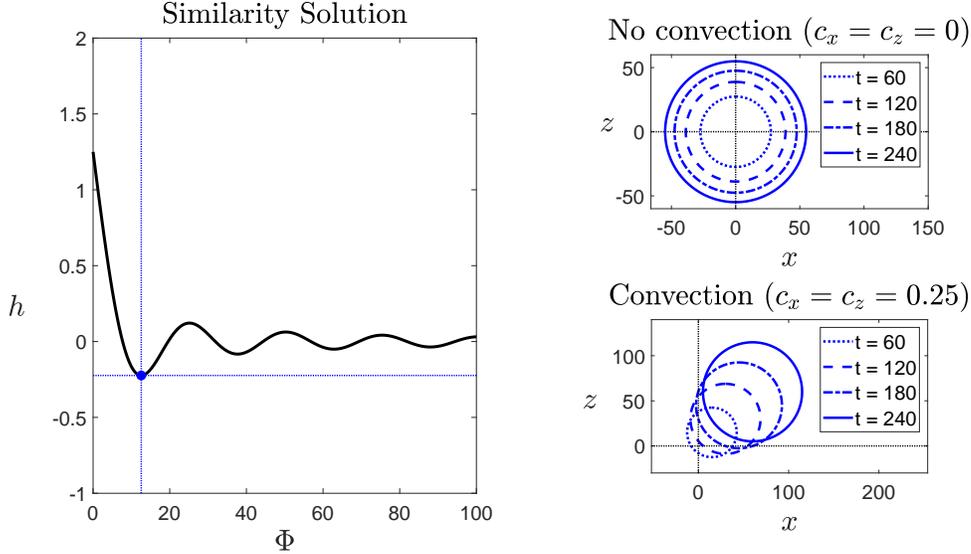}
\caption{
Similarity solution --Eq.~(\ref{eq:KP:WholeSolution})-- (left) with the specific point of $\Phi = 4B\pi$ chosen (\textcolor{blue}{$\bullet$}). This point is tracked through time for non convective (top right) and convective (bottom right) parameters. The disturbance is initiated at $(x,z)=(0,0)$ in accordance with Eq.~(\ref{eq:KP:LiteralOperator}). In both cases, the circles of constant $h$, given by Eqs.~(\ref{eq:KP:RingsOfConstantHeight}) and~(\ref{eq:KP:RingsOfConstantHeightNonconvective}), propagate out from the peak over time. For the convective case, the rings spread outwards more slowly than they convect. All data was generated with $A=10$ and $B=1$.
}
\label{fig:KP:SimilaritySolution}
\end{figure}

The similarity solution--Eq.~(\ref{eq:KP:Solution})-- can be used track the spatial behaviour of the response. For any given fixed value of $\Phi$, $h(\Phi)$ will be constant. Therefore, all values of $x$, $z$, and $t$ which result in the same $\Phi$ will be the same height. this also means that the apparent phase of the wave form is fixed for a given value of phi. Eq.~(\ref{eq:KP:PhiDef}) may be rewritten as:

\begin{equation}
\label{eq:KP:RingsOfConstantHeight}
R = \sqrt{(x-c_xt)^2
+(z-c_zt)^2} = \sqrt{\Phi t}.
\end{equation}
Thus, for any given height, the solution (fixed $\Phi$) moves radially relative to the peak as $\sqrt{t}$, but does so more slowly than the peak itself translates $(x=c_xt, z=c_zt)$.  Fig.~\ref{fig:KP:SimilaritySolution} provides a visualization of these features for a given value of $\Phi$.

\FloatBarrier

For the case where either $c_x$ or $c_z$ is nonzero, the peak itself convects away from its initializing location at $x=z=0$. From that perspective, the solution itself has a character akin to a convectively unstable system~\cite{HuerreRossi,huerre2000}, although here the maximum height of the system is constant. For this reason, we follow the convention of~\cite{barlow2011} in classifying these waves as ``convectively neutral."
It is worth noting that if both convective parameters are zero, then the expanding circles are all concentric, as shown in Fig.~\ref{fig:KP:SimilaritySolution}, and are given by

\begin{equation}
\label{eq:KP:RingsOfConstantHeightNonconvective}
\sqrt{x^2+z^2} = \sqrt{\Phi t}.
\end{equation}
In this case the system can said to be ``absolutely neutral" because the disturbance will eventually infect any given $(x,z)$ domain~\cite{barlow2011}. Unlike absolutely unstable systems, the solution here does not grow without bound, as the maximum value is the constant height of the peak.

\section{2D-CMH: 2D extension of algebraically growing 1D-CMH model}
\label{sec:NP:CMH}

\subsection{Problem statement}

We now consider the 2D extension of the 1D differential operator described in~\cite{Huber2020} and examine the response of that system. For reference in what follows, recall from Section~\ref{sec:General:Introduction} that the 1D and 2D problems are referred to as 1D-CMH and 2D-CMH, respectively. The well-posed system to examine the 2D response, $h$, is given by

 \begin{subequations}
 \label{eq:NP:OperatorFull}
\begin{equation}
\label{eq:NP:LiteralOperatorFull}
\Bigg[
\frac{\partial^2}{\partial t^2}
+\nabla^4
-\nabla^2\frac{\partial}{\partial t}
+B\frac{\partial}{\partial t}
+B^2
\Bigg]h=A\delta(x)\delta(z)\delta(t)
\end{equation}

\begin{equation}
\label{eq:NP:DelOperatorsFull}
\nabla^4 =
\frac{\partial^4}{\partial x^4}+
2\frac{\partial^4}{\partial x^2z^2}+
\frac{\partial^4}{\partial z^4},
\qquad
\nabla^2 =
\frac{\partial^2}{\partial x^2}+
\frac{\partial^2}{\partial z^2},
\end{equation}

\begin{equation}
\label{eq:NP:FullInitialConditions}
h(x,z,0)=0
,\qquad
\frac{\partial h}{\partial t}(x,z,0)=0
\end{equation}

\begin{equation}
\label{eq:NP:FullOtherConditions}
h\to 0
\quad\text{as}\quad
x,z\to\pm\infty
,\qquad
B\in \mathbb{R}.
\end{equation}
\end{subequations}
where $B$ is a real parameter affecting system stability (see~\cite{Huber2020}) and $A$ is a real forcing magnitude.
Note that Eq.~(\ref{eq:NP:OperatorFull}) is a simplified extension of the 1D-CMH operator found in~\cite{Huber2020}, as the terms responsible for convection have been removed.
As with the previous problem, zero valued initial conditions are chosen. Further justification for this choice is provided in Section~\ref{sec:NP:InitialConditions}.

\subsection{Classical stability analysis}

The classical stability analysis is performed by substituting Eq.~(\ref{eq:general:ModalForm}) into the homogeneous form of Eq.~(\ref{eq:NP:OperatorFull}) as described previously. A dispersion relation for $\omega(\textbf{k})$ is obtained that assures nontrivial solutions,

\begin{equation}
\label{eq:NP:DispersionRelationFull}
\omega=
\frac{1}{2}
\left(
-i(B+\eta)
\pm
\sqrt{
3\left(
	\eta-\frac{B}{3}
\right)^2
+\frac{8}{3}B^2
}
\right)
,\qquad
\eta = k_x^2+k_z^2.
\end{equation} 
Eq.~(\ref{eq:NP:DispersionRelationFull}) indicates that

\begin{equation}
\label{eq:NP:DispersionRelationImaginary}
\omega_i=-
\frac{(B+k_x^2+k_z^2)}{2}
\quad
\text{ and that }
\quad
\omega_{i,\mathrm{max}}=
-\frac{B}{2},
\end{equation}
and thus stability is determined by the sign of $B$ in Eq.~(\ref{eq:NP:DispersionRelationImaginary}).  According to the classical stability characterization --Eq.~(\ref{eq:general:ClassicalStabilityClassification})--, the system is unstable when $B<0$, stable when $B>0$, and neutrally stable when $B=0$. This is the same characterization given to the 1D-CMH system in~\cite{Huber2020}.  For $B=0$ (the situation examined in this paper) the dispersion relation is written for reference as

\begin{equation}
\label{eq:NP:DispersionRelationStable}
\omega=
\frac{1}{2}
\left(
-i\eta
\pm
\eta\sqrt{3}
\right)
,\quad
\eta =k_x^2+k_z^2,
\quad 
\eta\geq 0.
\end{equation} 
The corresponding operator is expressed as

\begin{subequations}
\label{eq:NP:Operator}
\begin{equation}
\label{eq:NP:LiteralOperator}
\Bigg[
\frac{\partial^2}{\partial t^2}
+\nabla^4
-\nabla^2\frac{\partial}{\partial t}
\Bigg]h=A\delta(x)\delta(z)\delta(t),
\end{equation}

\begin{equation}
\label{eq:NP:DelOperators}
\nabla^4 =
\frac{\partial^4}{\partial x^4}+
2\frac{\partial^4}{\partial x^2z^2}+
\frac{\partial^4}{\partial z^4},
\qquad
\nabla^2 =
\frac{\partial^2}{\partial x^2}+
\frac{\partial^2}{\partial z^2},
\end{equation}

\begin{equation}
\label{eq:NP:InitialConditions}
h(x,z,0)=0
,\qquad
\frac{\partial h}{\partial t}(x,z,0)=0,
\end{equation}

\begin{equation}
\label{eq:NP:OtherConditions}
h\to 0
\quad
\text{as}
\quad
x,z\to\pm\infty.
\end{equation}
\end{subequations}

\subsection{Initial Conditions}
\label{sec:NP:InitialConditions}
As discussed in the context of the 2D-KRK problem in Section~\ref{sec:KP:KRK}, care has been taken in choosing homogeneous constraints in Eqs.~(\ref{eq:NP:FullInitialConditions}) and~(\ref{eq:NP:InitialConditions}) so as to assure correct stability conclusions are drawn about the operator.  In particular, the effect of applying an impulse function to the initial surface velocity $\nicefrac{\partial h}{\partial t}$, is the same as applying the impulse forcing function included in Eq.~(\ref{eq:NP:LiteralOperator}) in an analogous way to the 2D-KRK problem of Section~\ref{sec:KP:KRK} (see Supplemental Material Section~\ref{sup:NP:InitialConditions}). Due to the complexity of the analysis method, the effect of a delta function in the initial surface height, $h(x,z,0)$ in Eq.~(\ref{eq:NP:InitialConditions}), was surveyed numerically via the Fourier Series Solution (FSS) provided in \ref{app:NP:FSS}. It should be noted that spatially infinite domain problems  require transforms to obtain solutions, but here a discrete Fourier series is utilized.  To do so, the domain of the operator is truncated to be sufficiently large such that changes in the domain length do not yield any changes in the response in the times examined~\cite{barlow2010}.
We find that the response from an initial surface height decays along all $V_x=\nicefrac{x}{t}$ and $V_z=\nicefrac{z}{t}$ velocities either at the same rate as, or faster than, the solution disturbed with only the surface velocity.
We conclude that an impulsive function forcing is a sufficient disturbance to extract the stability character of the operator. 
As a result, the analysis is conducted with just the impulse forcing function in Eq.~(\ref{eq:NP:LiteralOperator}) and the initial conditions given in Eq.~(\ref{eq:NP:InitialConditions}).
In what follows, the just-described FSS (\ref{app:NP:FSS}) is used to validate all asymptotic results in their regimes of validity and to generate full solutions to Eq.~(\ref{eq:NP:Operator}).
In the 3D plot to follow (Fig.~\ref{fig:NP:3DPlot}), the FSS is generated using 2000 terms each in $k_x$ and $k_z$. For the plots where only one direction is shown (Figs.~\ref{fig:NP:Dimensional_comparison} and~\ref{fig:NP:AsymptoticComparisonCompiled}), the 4000 terms were used in $k_x$ and  $k_z$ is not needed to generate solutions (because only $z=0$ was considered).

\subsection{Analysis}

The solution to Eq.~(\ref{eq:NP:Operator}) is found by taking Fourier transforms in $x$ and $z$ (resulting in the transformed variable $\hat{\hat{h}}_{xz}$) and the Laplace transform in $t$. The resulting Fourier inversion integral solution is given as

\begin{equation*}
h(x,z,t) = \frac{1}{4\pi^2}
	\int\limits_{-\infty}^{\infty} 
	\int\limits_{-\infty}^{\infty} 
	\hat{\hat{h}}_{xz}
	e^{ik_xx}
	e^{ik_zz}
	dk_xdk_z ,
\end{equation*}

\begin{equation}
\label{eq:NP:FourierSolution}
\hat{\hat{h}}_{xz}=
\Big(
Ae^{-\frac{\eta}{2}t}
\Big)
\frac{\sin
\left(\frac{\eta}{2}\sqrt{3} t
\right)}
{\frac{\eta}{2}\sqrt{3}}
,\qquad
\eta = k_x^2+k_z^2.
\end{equation}
Note that the result --Eq.~(\ref{eq:NP:FourierSolution})-- is similar in form to that of Eq.~(\ref{eq:KP:FourierSolution}) for the 2D-KRK problem~(Section~\ref{sec:KP:KRK}), with the significant difference being the $O(k_x^2,k_z^2)$ term in the damped exponential instead of the $O(k_x,k_z)$ term.  This difference in form leads to a significant increase in analysis complexity. 
In order to evaluate Eq.~(\ref{eq:NP:FourierSolution}), it is first converted to polar form as, $(k_x,k_z)\to(\xi,\theta)$, with radial component $\xi\in[0,\infty)$ and angular component $\theta\in[0,2\pi]$. The result is the new integral given by

\begin{equation*}
h(V_x,V_z,t) = 
\frac{A}{2\pi^2\sqrt{3}}
\int\limits_{0}^{\infty}
\frac{
\sin\left(
\frac{\sqrt{3}}{2}\xi^2t\right)
}{\xi}
e^{-\frac{\xi^2}{2}t}
\left(
\int\limits_{0}^{2\pi}
e^{i\mathcal{B}\xi 
t}d\theta
\right)
d\xi ,
\end{equation*}

\begin{equation}
\label{eq:NP:PolarForm}
\mathcal{B} = 
\left(
\cos(\theta)V_x
+\sin(\theta)V_z
\right)
,\qquad
V_x=
\frac{x}{t}
,\quad
V_z=
\frac{z}{t}.
\end{equation}
The inner integral in $\theta$ can be evaluated asymptotically through the method of steepest descent~(\ref{app:NP:ThetaIntegral}) in the limit as $t\to\infty$ holding $V_x$ and $V_z$ fixed. The result is 

\begin{equation}
\label{eq:NP:ThetaAsymptotic}
\int\limits_{0}^{2\pi}
e^{i\mathcal{B}\xi t}
d\theta\Bigg|_{\hat{V}\neq 0}\sim
2\cos
\left(
\xi\hat{V}t-\frac{\pi}{4}
\right)
\sqrt{\frac{2\pi}{\hat{V}\xi t}}
\quad \text{ as } t\to\infty.
\end{equation}
When $\hat{V}=0$, the $\theta$ integral can be evaluated exactly. The solution to the $\theta$ integral for this case is

\begin{equation}
\label{eq:NP:ThetaExact}
\int\limits_{0}^{2\pi}
e^{i\mathcal{B}\xi t}
d\theta\Bigg|_{\hat{V}=0}
=
\int\limits_{0}^{2\pi}
e^{0}
d\theta\Bigg|_{\hat{V}=0}
=2\pi.
\end{equation}
Eqs.~(\ref{eq:NP:ThetaAsymptotic}) and~(\ref{eq:NP:ThetaExact}) are substituted into Eq.~(\ref{eq:NP:PolarForm}) to yield

\begin{equation}
\label{eq:NP:XiIntegralAsymptotic}
h(\hat{V},t)\Big|_{\hat{V}\neq 0}\sim
\frac{A}{2\pi^2\sqrt{3}}
			\int\limits_0^{\infty}
			\frac{\sin
			\left(
			\frac{\sqrt{3}}{2}	
			\xi^2t
			\right)}{\xi}
			e^{-\frac{\xi^2}{2}t}
			\left(
			2\cos
			\left(
				\xi\hat{V}t
				-\frac{\pi}{4}
			\right)	\sqrt{\frac{2\pi}{\hat{V}\xi t}}
			\right)
			d\xi
			\qquad \text{ as } t\to\infty,
\end{equation}

\begin{equation}
\label{eq:NP:XiIntegralExact}
h(0,t) = 
\frac{A}{\pi\sqrt{3}}
\int\limits_{0}^{\infty}
\frac{
\sin\left(
\frac{\sqrt{3}}{2}\xi^2t\right)}
{\xi}
e^{-\frac{\xi^2}{2}t}
d\xi.
\end{equation}
Eq.~(\ref{eq:NP:XiIntegralExact}), which is exact, may be evaluated, through the variable substitution of $U=\frac{\sqrt{3}}{2}\xi^2t$, to obtain a constant height solution as

\begin{equation}
\label{eq:NP:V0Solution}
h\Big|_{\hat{V} = 0} = 
\frac{A}{6\sqrt{3}}
\text{ for all }
t.
\end{equation}
In the case where $\hat{V}\neq 0$, Eq.~(\ref{eq:NP:XiIntegralAsymptotic}) is rearranged through the substitution of $U= \xi^2t$ to obtain

\begin{equation}
\label{eq:NP:JIntegralCosine}
h \sim 
\frac{A}{2\pi^2\sqrt{3}}
\left(
	\frac{(2\pi)^{\frac{1}{2}}}
	{\hat{V}^{\frac{1}{2}}t^{\frac{1}{4}}}
\right)
\int\limits_0^{\infty}
\frac{\sin
	\left(
		\frac{\sqrt{3}}{2}
		U
	\right)
	}{
	U^{\frac{5}{4}}}
e^{-\frac{U}{2}}
	\cos
	\left(
		\hat{V}(Ut)^\frac{1}{2}
		-\frac{\pi}{4}
	\right)
dU
\qquad \text{ as } t\to\infty.
\end{equation}
To evaluate the integral in Eq.~(\ref{eq:NP:JIntegralCosine}), the cosine is interpreted as the real part of a complex integral as follows:

\begin{equation}
\label{eq:NP:HInTermsOfJ}
h \sim 
\frac{A}{2\pi^2\sqrt{3}}
\left(
	\frac{(2\pi)^{\frac{1}{2}}}
	{\hat{V}^{\frac{1}{2}}t^{\frac{1}{4}}}
\right)
Re\Bigg[
e^{
\left(
	-i\frac{\pi}{4}
\right)}
J
\Bigg]
\qquad \text{ as } t\to\infty,
\end{equation}

\begin{equation}
\label{eq:NP:JIntegral}
J = \int\limits_0^{\infty}
\frac{\sin
	\left(
		\frac{\sqrt{3}}{2}
		U
	\right)
	}{
	U^{\frac{5}{4}}}
e^{-\frac{U}{2}}
e^{
\left(
	i\hat{V}(Ut)^\frac{1}{2}
\right)}
dU.
\end{equation}
After performing the substitutions of $W=\sqrt{U}$ and $S = \sqrt{\hat{V}^2t}$, Eq.~(\ref{eq:NP:JIntegral}) can be written as

\begin{subequations}
\label{eq:NP:JIntegrals}
\begin{equation}
J = 2 ( J_1 + iJ_2 ),
\end{equation}

\begin{equation}
\label{eq:NP:JIntegral1}
J_1 = \int\limits_0^{\infty}
\frac{\sin
	\left(
		\frac{\sqrt{3}}{2}
		W^2
	\right)
	}{
	W^{\frac{3}{2}}}
e^{-\frac{W^2}{2}}
	\cos(WS)
dW,
\end{equation}

\begin{equation}
\label{eq:NP:JIntegral2}
J_2 = \int\limits_0^{\infty}
\frac{\sin
	\left(
		\frac{\sqrt{3}}{2}
		W^2
	\right)
	}{
	W^{\frac{3}{2}}}
e^{-\frac{W^2}{2}}
	\sin(WS)
dW.
\end{equation}

\end{subequations}

In the analysis of the integrals in Eqs.~(\ref{eq:NP:JIntegral1}) and~(\ref{eq:NP:JIntegral2}), traditional methods of asymptotic analysis fail as $S$ approaches infinity (which is consistent with the $t\to\infty$ limit taken in Eq.~(\ref{eq:NP:JIntegralCosine})). In particular, integration by parts and the method of steepest descent both cannot be used extract an asymptotic behaviour. A solution can, however, be found by complexifying the sines and evaluating the resulting integrals in terms of modified Bessel functions of the first kind, $\mathcal{I}_\nu(z)$. These can, in turn, be expanded asymptotically for large $S$. This is a simplified explanation for brevity; the full analysis can be found in Supplemental Material section \ref{sup:NP:JIntegral}.

The structure of the modified Bessel functions plays a key role in the evaluation of Eqs.~(\ref{eq:NP:JIntegral1}) and~(\ref{eq:NP:JIntegral2}) and is discussed here. The following asymptotic expansions were used in our analysis and are adjustments of those found in Abramowitz and Stegun~\cite{AbramowitzBessel} and Wolfram Alpha~\cite{WolframAlphaBessel}:

\begin{subequations}
\label{eq:NP:AllBesselInfo}
\begin{equation}
\label{eq:NP:BesselExpansion}
I_\nu(z) \sim
\frac{e^{z}}
{\sqrt{2\pi z}}
\mathbb{I}(\nu,z)
+
i\frac{e^{
-z+i\pi \nu(1-2m)-i\pi m
}}{\sqrt{2\pi z}}
\mathbb{J}(\nu,z)
\qquad
\text{ as } |z|\to\infty,
\end{equation}

\begin{equation}
m=\begin{cases}
0, &Im(z) > 0 \\
1, &Im(z) <0 
\end{cases}
,\qquad
\nu \in \mathbb{R}
,\qquad
z \in \mathbb{C}
,\qquad
z \notin \mathbb{R},
\end{equation}

\begin{multline}
\mathbb{I}[\nu,z]=
1
-\frac{(4\nu^2-1)}{8z}
+\frac{(4\nu^2-1)(4\nu^2-9)}
	{2!(8z)^2}
-\frac{(4\nu^2-1)(4\nu^2-9)(4\nu^2-25)}
	{3!(8z)^3}
\\
+\frac{(4\nu^2-1)(4\nu^2-9)(4\nu^2-25)(4\nu^2-49)}
	{4!(8z)^4}
+O\left(\frac{1}{z^5}\right),
\end{multline}

\begin{multline}
\mathbb{J}[\nu,z]=
1
+\frac{(4\nu^2-1)}{8z}
+\frac{(4\nu^2-1)(4\nu^2-9)}
	{2!(8z)^2}
+\frac{(4\nu^2-1)(4\nu^2-9)(4\nu^2-25)}
	{3!(8z)^3}
\\
+\frac{(4\nu^2-1)(4\nu^2-9)(4\nu^2-25)(4\nu^2-49)}
	{4!(8z)^4}
+O\left(\frac{1}{z^5}\right),
\end{multline}
\end{subequations}
these adjustments were necessary so that they agreed with the implementation of MATLAB's besseli() function over the full range of arguments required in our analysis.
Note that the only difference between $\mathbb{I}$ and $\mathbb{J}$ in Eq.~(\ref{eq:NP:AllBesselInfo}) is the sign of every other term.
An interesting issue encountered in the asymptotic analysis of Eq.~(\ref{eq:NP:JIntegral}) is the interactions between the leading order terms in Eq.~(\ref{eq:NP:BesselExpansion}).
If only the asymptotically dominant terms of Eq.~(\ref{eq:NP:BesselExpansion}) are considered, the following solution is constructed using the first five terms in $\mathbb{I}$ and $\mathbb{J}$,

\begin{equation}
J = b\sqrt{2\pi} 
\Bigg(
	C_1	S^{-\frac{3}{2}}
\\
	+C_2 S^{-\frac{7}{2}}	
\\
	+C_3 S^{-\frac{15}{2}}
\Bigg)
\qquad \text{ as }S\to\infty,
\end{equation}
where $C_1$, $C_2$,and $C_3$ are constants provided in Supplemental Material Section~\ref{sup:NP:OldLeadingOrder}. However, in accordance with Eq.~(\ref{eq:NP:HInTermsOfJ}), the solution for $J$ is multiplied by $e^{\nicefrac{-i\pi}{4}}$ and the real part is taken. When that is done, we find that

\begin{equation}
\label{eq:NP:CTermsZero}
Re\left[e^{-i\frac{\pi}{4}}C_1\right]=
Re\left[e^{-i\frac{\pi}{4}}C_2\right]=
Re\left[e^{-i\frac{\pi}{4}}C_3\right]=
Re\left[e^{-i\frac{\pi}{4}}J\right]=0.
\end{equation}
Thus, no asymptotic behaviour may be extracted from the leading order terms. Consequently, the sub-dominant terms of the Bessel expansion in Eq.~(\ref{eq:NP:AllBesselInfo}) must be included in the analysis of $J$, which leads to the the following result

\begin{multline}
\label{eq:NP:JAsymptoticSol}
J\sim
2\sqrt{2\pi}e^{i\frac{\pi}{4}}
\Bigg(
	\frac{1}{2}
	\sin\left(
		\frac{\sqrt{3}}{8} S^2
	\right)
	+\frac{\sqrt{3}}{2}
	\cos\left(
		\frac{\sqrt{3}}{8} S^2
	\right)
	-\frac{i}{2}
	\cos\left(
		\frac{\sqrt{3}}{8} S^2
	\right)
\\
	+\frac{i\sqrt{3}}{2}
	\sin\left(
		\frac{\sqrt{3}}{8} S^2
	\right)
\Bigg)
S^{-\frac{3}{2}}
e^{-\frac{S^2}{8}} 
+
	O\left(
		S^{-\frac{5}{2}}
		e^{-\frac{S^2}{8}}
	\right)
\quad
\text{ as } S\to\infty,
\end{multline}

Substituting Eq.~(\ref{eq:NP:JAsymptoticSol}) into Eq.~(\ref{eq:NP:HInTermsOfJ}) yields the asymptotic solution to Eq.~(\ref{eq:NP:Operator}). This solution, and Eq.~(\ref{eq:NP:V0Solution}), can be expressed as a function of a similarity variable $\Phi$ as

\begin{subequations}
\label{eq:NP:AsymptoticSolFull}
\begin{multline}
\label{eq:NP:Asymptotic}
h(\Phi)\Big|_{\Phi\neq 0} \sim 
\frac{A}{\pi\sqrt{3}}
\Bigg(
	\sin
	\left(
		\frac{\sqrt{3}}{8}\Phi
	\right)
\\
	+\sqrt{3}\cos
	\left(
		\frac{\sqrt{3}}{8}\Phi
	\right)
\Bigg)
\frac{e^{-\frac{\Phi}{8}}}
	{\Phi}
+
O\left(
	\Phi^{-\frac{3}{2}}
	e^{-\frac{\Phi}{8}}
\right)
\quad\text{ as } \Phi \to\infty,
\end{multline}

\begin{equation}
\label{eq:NP:V0SolutionPhi}
h\Big|_{\Phi = 0} = 
\frac{A}{6\sqrt{3}}
\text{ for all }
t,
\end{equation}

\begin{equation}
\Phi = \hat{V}^2t
,\qquad
\hat{V}=\sqrt{V_x^2+V_z^2}.
\end{equation}
\end{subequations}
Note that the similarity variable $\Phi$ is defined identically to that of the 2D-KRK problem --Eq.~(\ref{eq:KP:Solution})-- for $c_x=c_z=0$. However, unlike the solution to the 2D-KRK problem, Eq.~(\ref{eq:NP:Asymptotic}) is not exact. That said, the result in Eq.~(\ref{eq:NP:V0SolutionPhi}) is not subject to asymptotic constraints and is exact.

\begin{figure}[ht!]
\centering
\includegraphics[keepaspectratio,width=6in]{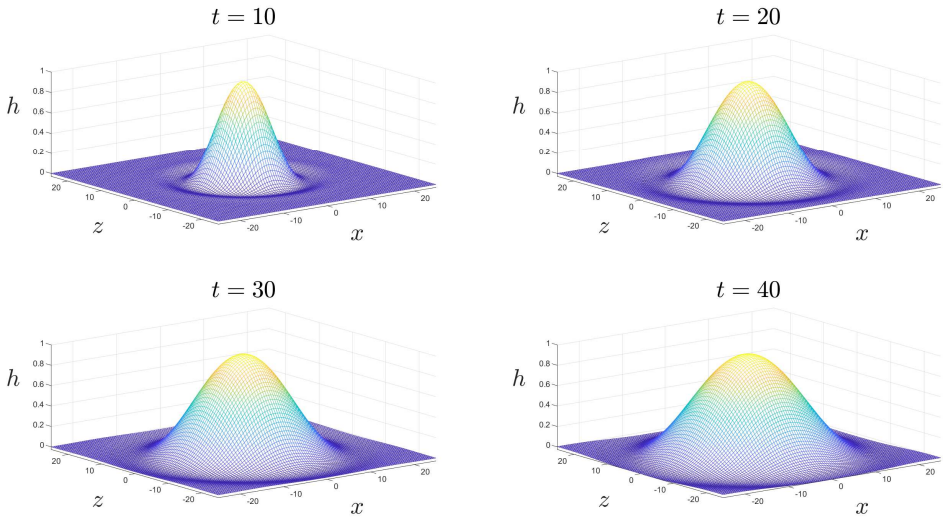}
\caption{
FSS to Eq.~(\ref{eq:NP:Operator}) for several different times. As time progresses, the peak expands radially and isotropically, but the maximum value remains constant. The data was generated using 2000 Fourier terms each in $k_x$ and $k_z$ and a value of $A=10$.
}
\label{fig:NP:3DPlot}
\end{figure}

\begin{figure}[ht!]
\centering
\includegraphics[keepaspectratio,width=6in]{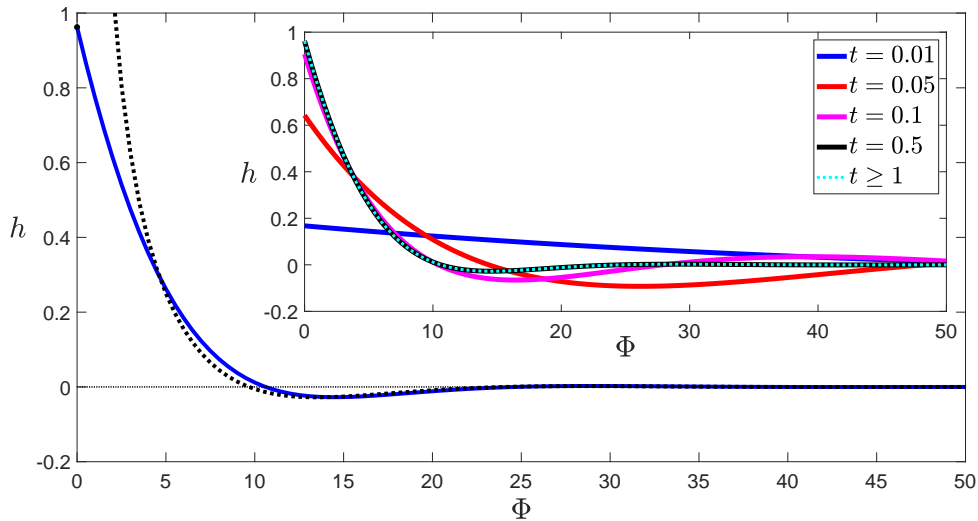}
\caption{
Comparison between the FSS to Eq.~(\ref{eq:NP:Operator}) (\ref{app:NP:FSS}) (solid blue line in main figure) and asymptotic solution --Eq.~(\ref{eq:NP:AsymptoticSolFull})-- (dashed black line in main figure) solutions for different values of $\Phi$. The two solutions agree at large values of $\Phi$ and at $\Phi=0$, but disagree at small values of $\Phi$. The $\Phi$-axis is marked with a dotted black line. The data in the main plot was generated for various $x$ and $z$ values $t=10$. The inset figure shows how the FSS (\ref{app:NP:FSS}) reaches a similarity solution as $t$ increases, well before the value of $t=10$ used in the main plot.
}
\label{fig:NP:AsymptoticComparisonCompiled}
\end{figure}

Similarly to the non-convective case for the 2D-KRK operator, the 2D-CMH solution has a single peak of constant height for all time, given exactly by Eq.~(\ref{eq:NP:V0Solution}) (or equivalently by Eq.~(\ref{eq:NP:V0SolutionPhi})). For all other velocities $\hat{V}$ in Eq.~(\ref{eq:NP:Asymptotic}), the solution decays exponentially.
The asymptotic solution given by Eq.~(\ref{eq:NP:Asymptotic}) is nonuniform in space and cannot resolve the behaviour for small $\Phi$ (see Fig.~\ref{fig:NP:AsymptoticComparisonCompiled}). For this reason, Fig.~\ref{fig:NP:3DPlot} uses the FSS (\ref{app:NP:FSS}) to construct the full response solution.

\FloatBarrier

\noindent
Notably in Fig.~\ref{fig:NP:3DPlot}, the response spreads radially in time; however the solution decays in time for every nonzero velocity (i.e. for a given $\hat{V}$). According to Eq.~(\ref{eq:NP:Asymptotic}), the asymptotic solution is constant for any given fixed value of $\Phi$; thus a given solution height, $h$, expands as $\sqrt{t}$ as given by Eq.~(\ref{eq:KP:RingsOfConstantHeightNonconvective}).
Because convective terms in the operator are zero as discussed above, the response is symmetric about the origin (the location of the initiating disturbance).      As the peak at $x=z=0$ does not grow or decay, the result may be considered ``absolutely neutral" as described for 2D-KRK in Section~\ref{sec:KP:SpatialStability}.

\FloatBarrier
As mentioned previously, the asymptotic solution --Eq.~(\ref{eq:NP:Asymptotic})-- cannot accurately predict the solution response for small values of $\Phi$, where all of the nontrivial behavior occurs.  Nevertheless, the behavior of the asymptotic solution is useful to examine key differences between  the 1D and 2D response propagation to follow.
Below a value of approximately $\Phi=11.4$, the asymptotic solution and the FSS begin to disagree, based on a criterion that the absolute difference be less than $1\%$ of the peak height (see~\ref{app:NP:DivergenceCriterion}).
It is worth noting that, for sufficiently long time, the FSS (\ref{app:NP:FSS}) tends towards a function solely of the similarity variable $\Phi = \hat{V}^2t$, as shown in the insert of Fig.~\ref{fig:NP:AsymptoticComparisonCompiled}.
This allows for a compact comparison shown in the main window of Fig.~\ref{fig:NP:AsymptoticComparisonCompiled} between the Fourier series and the the long-time asymptotic expansion, the latter of which is solely a function of $\Phi$ for all time, but only accurate for long-time and fixed $\hat{V}$; this is shown by the agreement of the two solutions in Fig.~\ref{fig:NP:AsymptoticComparisonCompiled} as $\Phi\to\infty$.

\section{Discussion}
\label{sec:General:Discussion}

\subsection{Comparison between 1D-KRK and 2D-KRK}
	\label{sec:KP:DimensionalComparison}

We now compare and contrast the 2D-KRK response propagation determined above in Section~\ref{sec:KP:KRK} with the 1D-KRK response from previous work. To enable a direct comparison, we write the 1D solution from~\cite{king2016} with only a forcing disturbance as

\begin{subequations}
\begin{multline}
\label{eq:KP:1DSolution}
h(x,z) = 
t\frac{\left|V_x-c_x\right|}{2B}
\left[
	S\left(
		\frac{\left|V_x-c_x\right|}{2B^{\frac{1}{2}}}
		t^{\frac{1}{2}}
	\right)
	-
	C\left(
		\frac{\left|V_x-c_x\right|}{2B^{\frac{1}{2}}}
		t^{\frac{1}{2}}
	\right)
\right]
\left[
	A
\right]
\\+
\frac{1}{\sqrt{\pi}}
\cos\left[
	\frac{\pi}{4}
	-\frac{(V_x-c_x)^2t}{4B}
\right]
\left(
	\frac{A}{\sqrt{B}}t^{\frac{1}{2}}
\right),
\end{multline}
where the Fresnel integrals $S(z)$ and $C(z)$ are defined as

\begin{equation}
S(z)=
\sqrt{\frac{2}{\pi}}
\int\limits_0^z
\sin(k^2)dk,
\qquad
C(z)=
\sqrt{\frac{2}{\pi}}
\int\limits_0^z
\cos(k^2)dk.
\end{equation} 
\end{subequations}
Note that, along the peak ($V_x=c_x$), Eq.~(\ref{eq:KP:1DSolution}) becomes
\begin{equation}
\label{eq:KP:1DSolutionV0}
h(x,z)\Big|_{V_x=c_x} = 
\frac{A}{\sqrt{2B\pi}}
t^{\frac{1}{2}}.
\end{equation}
The asymptotic behaviour of Eq.~(\ref{eq:KP:1DSolution}) for $V_x\neq c_x$ is provided in~\cite{king2016} (under the simplifications used above) as

\begin{equation}
\label{eq:KP:1DAsymptotic}
h(x,t)\big|_{V_x\neq c_x}\sim
\left(
	\frac{AB^{\frac{1}{2}}2^{\frac{3}{2}}}
	{\sqrt{2\pi}|V_x-c_x|^{\frac{5}{2}}}
	t^{-\frac{1}{2}}
\right)
\cos\left[
	\left(
		\frac{|V_x-c_x|}{2}
	\right)^2
	\frac{t}{B}
	+\frac{\pi}{4}
\right]
\\
+O(t^{-1})
\quad
\text{ as } t\to\infty.
\end{equation}

\begin{figure}[ht!]
\centering
\includegraphics[keepaspectratio,width=6in]{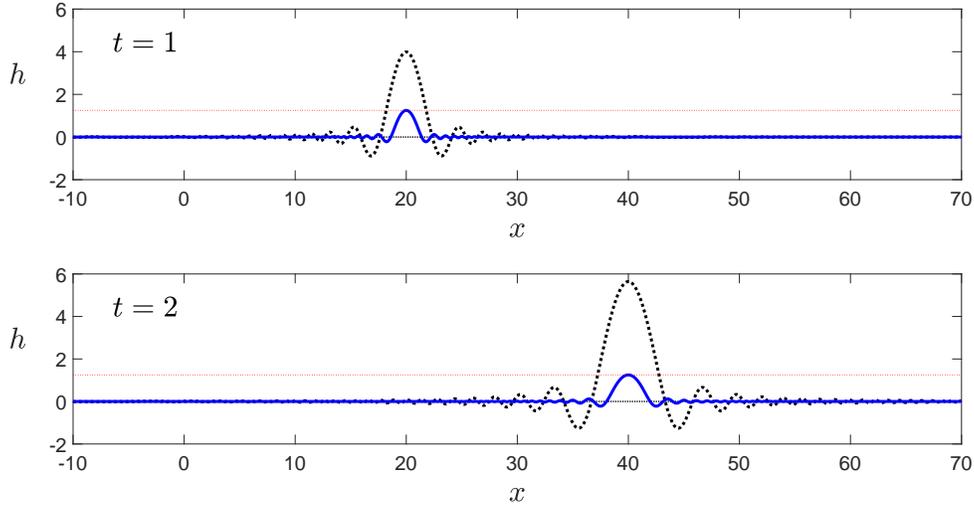}
\caption{
Comparison between the convective solutions of 1D-KRK (dotted black curve) and 2D-KRK (Solid blue curve) for $t=1$ and $t=2$. Short times are chosen to capture the behaviour of 1D-KRK before the algebraic growth swamps out the constant height of 2D-KRK.
Red horizontal lines mark the height of the 2D peak, which remains constant. 
The 1D-KRK response is generated using Eq.~(\ref{eq:KP:1DSolution}), and the 2D-KRK response is generated using Eq.~(\ref{eq:KP:WholeSolution}), both with $A=10$ and $B=1$.
}
\label{fig:KP:Dimensional_comparison_convection}
\end{figure}

\begin{figure}[ht!]
\centering
\includegraphics[keepaspectratio,width=6in]{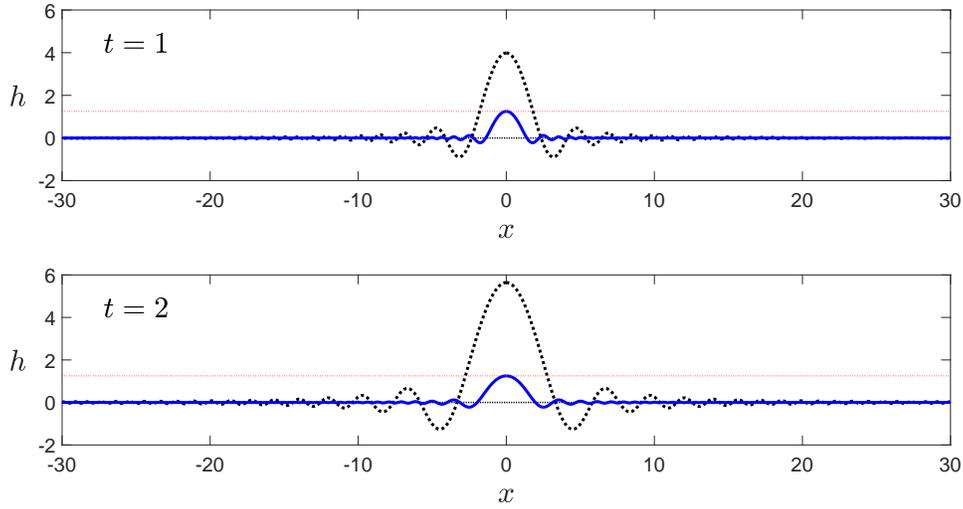}
\caption{
Comparison between the non-convective ($c=0$ for 1D, $c_x=c_z=0$ for 2D) solutions of 1D-KRK (dotted black curve) and 2D-KRK (Solid blue curve) for $t=1$ and $t=2$. Short times are chosen to capture the response features of 1D-KRK before the growing peak makes the axis scale too large to see the solution features of the 2D-KRK.
Red horizontal lines mark the height of the 2D peak, which remains constant. 
The 1D-KRK response is generated using Eq.~(\ref{eq:KP:1DSolution}), and the 2D-KRK response is generated using Eq.~(\ref{eq:KP:WholeSolution}), both with $A=10$ and $B=1$.
}
\label{fig:KP:Dimensional_comparison_no_convection}
\end{figure}

A significant distinction between the 1D-KRK and 2D-KRK solutions is that the peak of the former grows in time in accordance with Eq.~(\ref{eq:KP:1DSolutionV0}) while the peak of the latter stays constant. Additionally, the 1D-KRK
solution response has nontrivial features that extend farther (i.e., larger breadth) than those of the 2D-KRK solution (see Figs.~\ref{fig:KP:Dimensional_comparison_convection} and~\ref{fig:KP:Dimensional_comparison_no_convection}). The difference is further demonstrated by plotting the data sets normalized by their peaks (see~\ref{app:KP:PeakWidth} for plot).

\FloatBarrier

\FloatBarrier

\begin{figure}[ht!]
\centering
\includegraphics[keepaspectratio,width=7in]{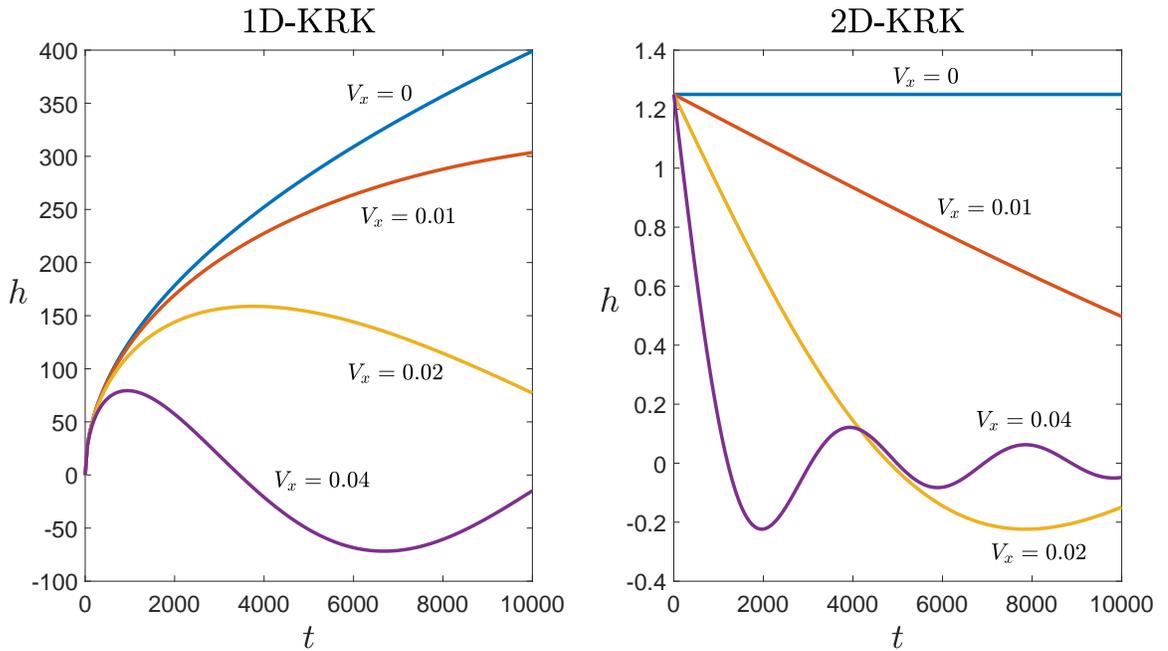}
\caption{
Non-convective results for small velocities (i.e. fixed velocities $V_x\equiv\nicefrac{x}{t}$) for the 1D- and 2D-KRK. In the 1D problem (left), the solution will grow transiently along non-zero velocities, but will ultimately decay in time. In the 2D problem (right), the solution immediately starts decaying along all non-zero velocities. The data for both cases were generated with the impulsive forcing amplitude of $A=10$, the parameter value of $B=1$, and the underlying velocities of $c_x = c_z = 0$. The 1D-KRK response is generated using the exact solution --Eq.~(\ref{eq:KP:1DSolution})-- adapted from~\cite{king2016} with $c=0$. The 2D solution is generated using Eq.~(\ref{eq:KP:WholeSolution}).
}
\label{fig:KP:Near_zero_comparison}
\end{figure}

The difference in the width of the 1D and 2D responses can be partially explained by the fact that the 1D solution exhibits transient growth (Fig.~\ref{fig:KP:Near_zero_comparison}), but the 2D response exhibits no such behaviour. Every nonzero $\hat{V}$ value in the 2D case immediately starts decaying,  as evidenced by the derivative of $h$ in Eq.~(\ref{eq:KP:Solution}) with respect to $t$,

\begin{equation}
\label{eq:KP:TimeDerivative}
\frac{\partial h}{\partial t}\bigg|_{t=0}=
-\frac{A\hat{V}^2}{16B^2\pi}.
\end{equation}
Past the transience, the solutions for every nonzero $\hat{V}$ decay faster in the 2D case (i.e. $O(\nicefrac{1}{t})$ from Eq.~(\ref{eq:KP:PeakSolution})) than in the 1D case (i.e. $O(t^{\nicefrac{-1}{2}})$ from Eq.~(\ref{eq:KP:1DSolutionV0})), further curtailing the peak width.
In both cases, the convective term does not change the shape of the disturbance, it only translates it in a given direction. For the 2D solution, this can be seen explicitly in the similarity solution --Eq.~(\ref{eq:KP:Solution})--, where $h$ is a function of the velocity relative to the traveling peak rather than that relative to the origin.

\FloatBarrier

\noindent
Overall, we see that the 1D-KRK response grows algebraically in Eq.~(\ref{eq:KP:1DSolutionV0}) compared to the constant height of the 2D-KRK peak in Eq.~(\ref{eq:KP:PeakSolution}) and this reduction in amplitude by a factor of $t^{\nicefrac{1}{2}}$ carries over to other fixed velocities. In particular, as noted above, the 1D-KRK response goes as $t^{\nicefrac{-1}{2}}$ from Eq.~(\ref{eq:KP:1DAsymptotic}) and in 2D-KRK, the response goes as $\nicefrac{1}{t}$ from Eq.~(\ref{eq:KP:Asymptotic}).

\subsection{Comparison between 1D-CMH and 2D-CMH}
	\label{sec:NP:Dimesional comparison}
\begin{figure}[ht!]
\centering
\includegraphics[keepaspectratio,width=7in]{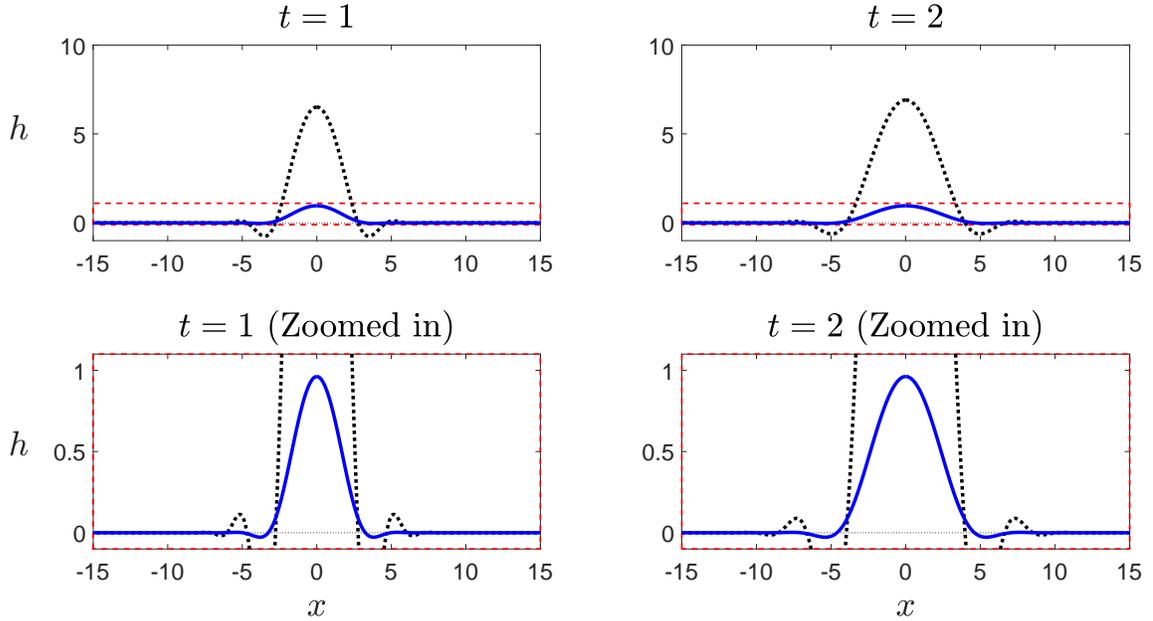}
\caption{
Comparison between the 1D solution (dotted black line) of~\cite{Huber2020} and the 2D solution --Eq.~(\ref{eq:NP:Operator})-- (solid blue line). Both solutions are generated using the FSS (\ref{app:NP:FSS}) with $A = 10$. The top row is zoomed out to show both solutions. The bottom row row) is zoomed in on the red dashed region to show that the 2D solution has the same structure as the 1D solution, just on a smaller scale.
}
\label{fig:NP:Dimensional_comparison}
\end{figure}

We now compare and contrast the 2D-CMH response propagation determined in Section~\ref{sec:NP:CMH} with the 1D-CMH responses from previous work. To enable comparison, we write write the 1D solution from~\cite{Huber2020} without convective terms and with only a forcing disturbance as

\begin{equation}
\label{eq:NP:1DSolution}
h(x,t)\big|_{V_x\neq 0}\sim
\frac{e^{-\frac{V_x^2}{8}t}}
{\sqrt{3\pi t}}
\left(
	\frac{4A}{V_x^2}
\right)
\cos\left[
\frac{\sqrt{3}V_x^2t}{8}
\right]
\quad
\text{ as } t\to\infty.
\end{equation}
The behaviour of Eq.~(\ref{eq:NP:1DSolution}) for $V_x=0$ is given exactly in~\cite{Huber2020} as

\begin{equation}
\label{eq:NP:1DSolutionV0}
h(x,t)\big|_{V_x=0}=A
\sqrt{\frac{t}{3\pi}}.
\end{equation}
A significant distinction between the 1D-CMH and 2D-CMH solutions is that the peak of the former grows in time in accordance with Eq.~(\ref{eq:NP:1DSolutionV0}) while the peak of the latter stays constant in accordance with Eq.~(\ref{eq:NP:V0Solution}). Fig.~\ref{fig:NP:Dimensional_comparison} provides a comparison of the 1D-CMH and 1D-CMH solutions obtained via the Fourier Series Solution (FSS) provided in \ref{app:NP:FSS}.

\FloatBarrier

As with the previous operator, there is a large difference in amplitude between the 1D and 2D cases. At the velocity of the peak, $\hat{V}=0$, the 1D-CMH solution grows algebraically as $t^{\nicefrac{1}{2}}$ from Eq.~(\ref{eq:NP:1DSolutionV0}) and the 2D-CMH solution is a constant from Eq.~(\ref{eq:NP:V0Solution}). For nonzero $\hat{V}$, the increase in dimension leads to a decay rate in 2D-CMH of the form $e^{-t}(t^{-1})$ from Eq.~(\ref{eq:NP:Asymptotic}) which is a factor of $t^{\nicefrac{-1}{2}}$ faster than the decay rate in 1D-CMH  of $e^{-t}(t^{\nicefrac{-1}{2}})$ from Eq.~(\ref{eq:NP:1DSolution}). Unlike the behaviour shown in figure~\ref{fig:KP:Dimensional_comparison_no_convection}, both cases here have the same controlling factor of $e^{-t}$. As such, there is not a noticeable difference between the widths of the responses.

\section{Conclusions}
\label{sec:General:Conclusions}

The linear operators studied by King et al.~\cite{king2016} and Huber et al.~\cite{Huber2020} have been extended to 2D through an identical methodology; time  derivatives are kept the same, and spatial derivatives are replaced with 2D del operators.  For both operators, we find that an increase in the dimensionality leads to a decrease in growth rates and an increase in decay rates. The changes are equivalent for both operators, and scale the response character by a factor of $t^{\nicefrac{-1}{2}}$.
The algebraic growth of $t^{\nicefrac{1}{2}}$ at the peak velocities in both~\cite{king2016}
and in~\cite{Huber2020} is reduced to a constant height proportional to $t^0$. Along all the other velocities, the behaviour is reduced from $(t^{\nicefrac{-1}{2}})$ and $(e^{\nicefrac{-\hat{V}}{8}}t^{\nicefrac{-1}{2}})$ to $(t^{-1})$ and $(e^{\nicefrac{-\hat{V}}{8}}t^{-1})$ respectively. 
The propagation of wave features in 2-D expand as $t^{\nicefrac{1}{2}}$, which is slightly more slowly than the corresponding 1D case for the KRK operator.
Both solutions can be expressed in terms of the similarity variable $\Phi$ (in 2D-KRK, for all time; in 2D-CMH, only for large time) which couples time and the radial velocity relative to the peak and allows for relevant features of the solutions to be extracted.

\section{Bibliography}
\bibliographystyle{unsrt}
\bibliography{neutral}

\begin{thebibliography}{10}

\bibitem{ibrahim1995}
EA~Ibrahim.
\newblock Spatial instability of a viscous liquid sheet.
\newblock {\em Journal of Propulsion and Power}, 11(1):146--152, 1995.

\bibitem{lin2003}
Sung Lin.
\newblock {\em Breakup of liquid sheets and jets}.
\newblock Cambridge University Press, 2003.

\bibitem{el2015}
Mohamed~F El-Sayed, GM~Moatimid, FMF Elsabaa, and MFE Amer.
\newblock Hydrodynamic instabilities of two viscoelastic liquid sheet models in
  an inviscid gas medium.
\newblock {\em Atomization and Sprays}, 25(2), 2015.

\bibitem{paul2004}
Edward~L Paul, Victor~A Atiemo-Obeng, and Suzanne~M Kresta.
\newblock {\em Handbook of industrial mixing: science and practice}.
\newblock John Wiley \& Sons, 2004.

\bibitem{cohen1992}
Edward Cohen and Edgar Gutoff.
\newblock {\em Modern Coating and Drying Technology}.
\newblock Wiley, 1992.

\bibitem{kistler1997}
Stephan~F Kistler and Peter~M Schweizer.
\newblock {\em Liquid film coating}.
\newblock Springer, 1997.

\bibitem{weinstein2004}
Steven~J Weinstein and Kenneth~J Ruschak.
\newblock Coating flows.
\newblock {\em Annu. Rev. Fluid Mech.}, 36:29--53, 2004.

\bibitem{rayleigh1880}
Lord Rayleigh.
\newblock On the stability of certain fluid motions.
\newblock {\em Proc. Math. Soc. Lond.}, 11:57--70, 1880.

\bibitem{Chandrasekhar}
S.~Chandrasekhar.
\newblock {\em Hydrodynamic and Hydromagnetic Stability}.
\newblock Clarendon Press, 1968.

\bibitem{HuerreRossi}
P.~Huerre and M.~Rossi.
\newblock Hydrodynamic instabilities in open flows.
\newblock In C.~Godr{\`e}che and P.~Manneville, editors, {\em Hydrodynamics and
  Nonlinear Instabilities}, pages 81--288. Cambridge University Press, 1998.

\bibitem{huerre2000}
P~Huerre.
\newblock {\em Perspectives in Fluid Dynamics}, chapter Open shear flow
  instabilities, pages 159--229.
\newblock Cambridge University Press, 2000.

\bibitem{king2016}
K.~R. King, S.~J. Weinstein, P.~M. Zaretzky, M.~Cromer, and N.~S. Barlow.
\newblock Stability of algebraically unstable dispersive flows.
\newblock {\em Phys. Rev. Fluids}, 1(073604), 2016.

\bibitem{Huber2020}
Colin Huber, Meaghan Hoitt, Nathaniel~S Barlow, Nicole Hill, Kimberlee
  Keithley, and Steven~J Weinstein.
\newblock {On the stability of waves in classically neutral flows}.
\newblock {\em IMA Journal of Applied Mathematics}, 85(2):309--340, 04 2020.

\bibitem{whitham2011}
Gerald~Beresford Whitham.
\newblock {\em Linear and nonlinear waves}, volume~42.
\newblock John Wiley \& Sons, 2011.

\bibitem{lighthill2001}
James Lighthill.
\newblock {\em Waves in fluids}.
\newblock Cambridge university press, 2001.

\bibitem{barlow2011}
N.~S. Barlow, B.~T. Helenbrook, and S.~P. Lin.
\newblock Transience to instability in a liquid sheet.
\newblock {\em J. Fluid. Mech.}, 666:358--390, 2011.

\bibitem{barlow2010}
N.~S. Barlow, B.~T. Helenbrook, S.~P. Lin, and S.~J. Weinstein.
\newblock An interpretation of absolutely and convectively unstable waves using
  series solutions.
\newblock {\em Wave Motion}, 47(8):564--582, 2010.

\bibitem{AbramowitzBessel}
M.~Abramowitz and I~Stegun.
\newblock {\em Handbook of Mathematical Functions}, page 377.
\newblock Dover, 1972.

\bibitem{WolframAlphaBessel}
Wolfram$|$Alpha.
\newblock
  \url{https://www.wolframalpha.com/input?i=expansion+of+modified+bessel+of+the+first+kind+as+z+goes+to+infinity}.

\bibitem{bender1999}
Carl~M Bender and Steven~A Orszag.
\newblock {\em Advanced Mathematical Methods for Scientists and Engineers I}.
\newblock Springer Science \& Business Media, 1999.

\bibitem{WolframJ1Integral}
Wolfram$|$Alpha.
\newblock
  \url{https://www.wolframalpha.com/input?i=integrate+x%5E%28-1%2F2%29*e%5E%28-b*x%5E2%29*sin%28c*x%29+from+0+to+infinity%2C+real%28b%29%3E0}.

\bibitem{WolframJ2Integral}
Wolfram$|$Alpha.
\newblock
  \url{https://www.wolframalpha.com/input?i=integrate+x%5E%28-3%2F2%29*e%5E%28-b*x%5E2%29*sin%28c*x%29+from+0+to+infinity%2C+real%28b%29%3E0}.

\end{thebibliography}

\begin{appendix}

\section{2D-KRK: Analysis}
	\label{app:KP:KRK}
\subsection{Effect of initial conditions on solution response}
\label{app:KP:InitialConditions}

Working from the 2D-KRK operator --Eq.~(\ref{eq:KP:Operator})-- with the forcing amplitude set to zero, we obtain

\begin{multline}
\label{eq:app:KP:InitalConditionsEquation}
\left(\frac{\partial^2h}{\partial t^2}
+2c_x\frac{\partial^2h}{\partial t\partial x}
+2c_z\frac{\partial^2h}{\partial t\partial z}
+c_x^2\frac{\partial^2h}{\partial x^2}
+2c_xc_z\frac{\partial^2h}{\partial x \partial z}
+c_z^2\frac{\partial^2h}{\partial z^2}\right)
\\+
B^2
\left(
\frac{\partial^4h}{\partial x^4}
+2\frac{\partial^4h}{\partial x^2 \partial z^2}
+\frac{\partial^4h}{\partial z^4}
\right)=0.
\end{multline}
Instead of the homogeneous initial conditions in Eq.~(\ref{eq:KP:InitialConditions}), we apply

\begin{equation}
h(x,z,0)=H_0\delta(x)\delta(z),
\quad
\frac{\partial h}{\partial t}(x,z,0)=U_0\delta(x)\delta(z),
\end{equation}
where $h\to 0$ as $x,z\to\pm\infty$. Taking the Fourier transforms of Eq.~(\ref{eq:app:KP:InitalConditionsEquation}) in the $x$ and $z$ directions, we obtain

\begin{multline}
\label{eq:app:KP:ODE}
\frac{d^2\hat{\hat{h}}_{xz}}{d t^2}
+2i c_x k_x\frac{d\hat{\hat{h}}_{xz}}{d t}
+2i c_z k_z\frac{d\hat{\hat{h}}_{xz}}{d t}
-c_x^2 k_x^2\hat{\hat{h}}_{xz}
-2 c_x c_z k_x k_z\hat{\hat{h}}_{xz}
-c_z^2 k_z^2\hat{\hat{h}}_{xz}
\\+B^2k_x^4\hat{\hat{h}}_{xz}
+2B^2 k_x^2 k_z^2\hat{\hat{h}}_{xz}
+B^2 k_z^4\hat{\hat{h}}_{xz}
=0,
\end{multline}
where

\begin{equation}
\label{eq:app:KP:ODEConditions}
\hat{\hat{h}}_{xz}(0)=H_0,
\quad
\frac{d\hat{\hat{h}}_{xz}}{dt}(0)=U_0,
\end{equation}

\begin{equation}
\psi = c_xk_x + c_zk_z,
\quad
\eta = B(k_x^2+k_z^2).
\end{equation}
The solution to the 2nd order ordinary differential equation --Eq.~(\ref{eq:app:KP:ODE})-- subject to Eq.~(\ref{eq:app:KP:ODEConditions}) is

\begin{equation}
\label{eq:app:KP:hHatHat}
\hat{\hat{h}}_{xz}= 
e^{-i\psi t}
\left(
\frac{U_0+(i\psi + i\eta)H_0}
{\eta}
\right)
\left( 
\frac{
	e^{i\eta t}
	-
	e^{-i\eta t}
	}{2i}
\right)
+H_0 
e^{-i\psi t}
e^{-i\eta t}.
\end{equation}
The forced solution --Eq.(\ref{eq:KP:Solution})-- is superimposed on Eq.~(\ref{eq:app:KP:hHatHat}) to obtain

\begin{equation}
\label{eq:app:KP:TotalFourierSolution}
\hat{\hat{h}}_{xz,Total}= 
e^{-i\psi t}
\left(
\frac{(A+U_0)+(i\psi + i\eta)H_0}
{\eta}
\right)
\left( 
\frac{
	e^{i\eta t}
	-
	e^{-i\eta t}
	}{2i}
\right)
+H_0 
e^{-i\psi t}
e^{-i\eta t}.
\end{equation}
By inspection of Eq.~(\ref{eq:app:KP:TotalFourierSolution}), the effect of the forcing magnitude $A$ and the initial surface velocity $U_0$ are equivalent. Therefore, $U_0$ can be taken to be zero without loss of response character.

To examine the effect of a initial surface height, the values of $A$ and $U_0$ are set to zero in Eq.~(\ref{eq:app:KP:TotalFourierSolution}) and the inversion integrals are utilized to yield

\begin{equation*}
h = H_0
\int\limits_{-\infty}^{\infty}
\int\limits_{-\infty}^{\infty}
\left(
\left(
\frac{\psi + \eta}
{\eta}
\right)
\left( 
\frac{
	e^{i\eta t}
	-
	e^{-i\eta t}
	}{2}
\right)
+e^{-i\eta t}
\right)
e^{-i\psi t}
e^{ik_xV_xt}
e^{ik_zV_zt}
dk_zdk_x,
\end{equation*}

\begin{equation}
\label{eq:app:KP:H0Integral}
\psi = c_xk_x + c_zk_z,
\quad
\eta = k_x^2+k_z^2,
\quad
V_x = \frac{x}{t},
\quad
V_z = \frac{z}{t}.
\end{equation}
Through the steps included in Supplemental Material Section~\ref{sup:KP:InitialConditions}, Eq.~(\ref{eq:app:KP:H0Integral}) is solved exactly to yield
 
\begin{equation*}
h(x,z,t)=
\frac{H_0}{4Bt\pi}
\Bigg(
1
-2
\frac{c_x(V_x-c_x)+c_z(V_z-c_z)}{\hat{V}}
\Bigg)
\sin
\left(\frac{\hat{V}t}{4B}\right),
\end{equation*}

\begin{equation}
\label{eq:app:KP:H0Sol}
\hat{V} = \sqrt{(V_x-c_x)^2+(V_z-c_z)^2}.
\end{equation}
Note that $h$ is not localized-- it exists at non-negligible amplitude at all values of $\hat{V}$ for any value of $t$. This violates the condition of $h\to 0$ as $x,z\to\infty$. This solution is admitted because the Fourier transform of the integral exists due to the rapid oscillations in the integrand.

	\subsection{Determination of $\mathcal{Q}$ and $\mathcal{W}$ in Eq.~(\ref{eq:KP:QWSplit}) via contour integration}
		\label{app:KP:ContourIntegration}
	
From Eq.~(\ref{eq:KP:QWSplit}), the integrals $\mathcal{Q}$ and $\mathcal{W}$ may be expressed in terms of separate $x$ and $z$ integrals as

\begin{equation}
\label{eq:app:KP:QWForms}
\mathcal{Q}=\mathcal{Q}_x\mathcal{Q}_z,
\quad
\mathcal{W}=\mathcal{W}_x\mathcal{W}_z.
\end{equation}

\begin{figure}[ht!]
\centering
\includegraphics[keepaspectratio,width=4in]{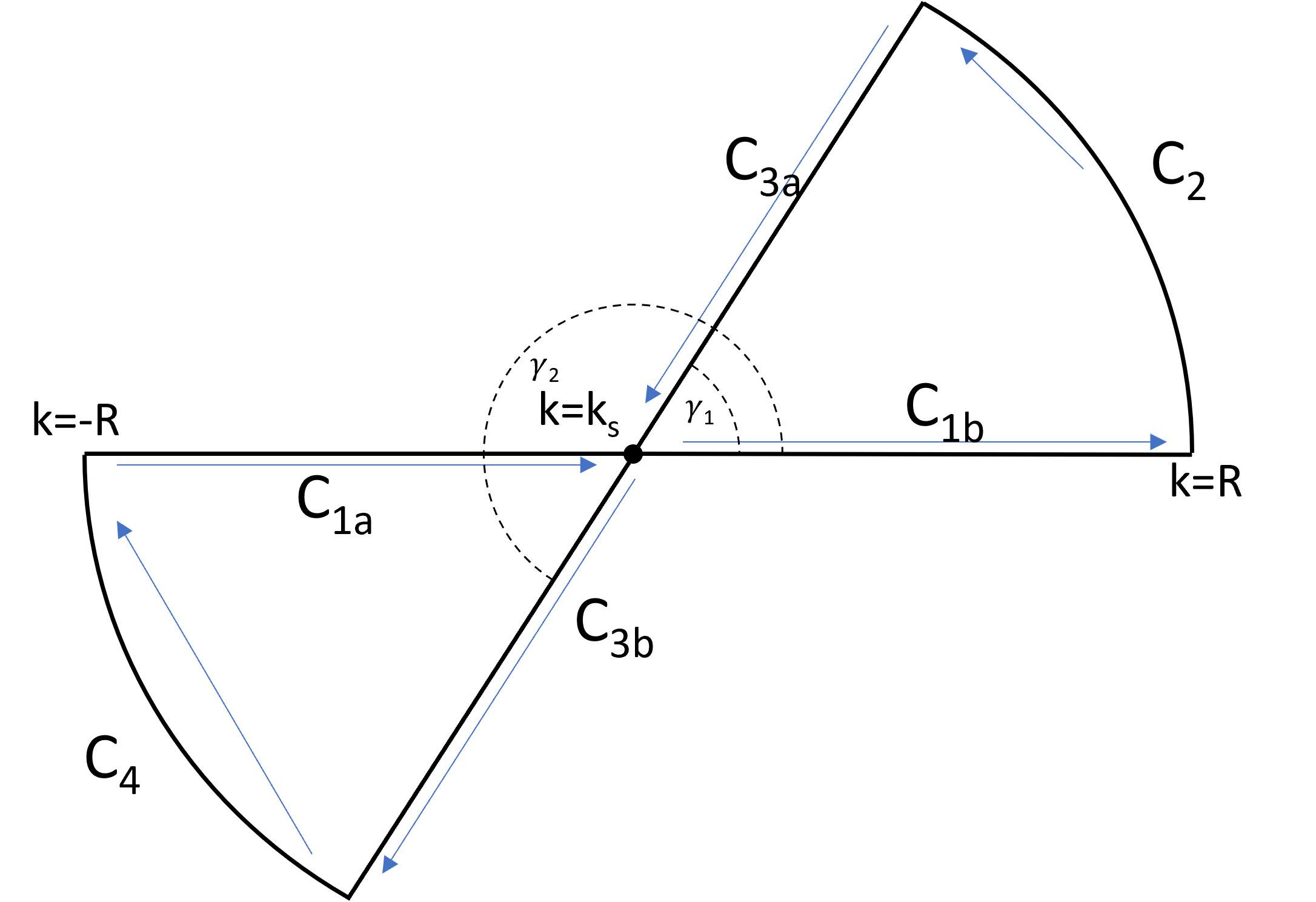}
\caption{Integration path for $\mathcal{Q}_x$. Note that Eq.~(\ref{eq:app:KP:SaddlePoint}) indicates that the saddle point $k_s$ always lies on the real axis.}
\label{fig:app:KP:SampleContour}
\end{figure}

In Eq.~(\ref{eq:app:KP:QWForms}), the integral subscripts $x$ and $z$ correspond respectively to the integrated variables $k_x$ and $k_z$ in Eq.~(\ref{eq:KP:QWSplit}); the definitions of $\mathcal{Q}_x$, $\mathcal{Q}_z$, $\mathcal{W}_x$, and $\mathcal{W}_z$ can thus be made by inspection of Eq.~(\ref{eq:KP:QWSplit}). 
The analysis for $\mathcal{Q}_x$ is presented below; the analysis of the remaining sub-integrals closely follows that provided here, and are relegated to Supplemental Material section \ref{sup:KP:QWIntegrals}.
To begin, we write
\begin{equation}
\label{eq:app:KP:SampleIntegral}
\mathcal{Q}_x=
\int\limits_{-\infty}^{\infty}
e^{i\left(\Phi(k_x) \right)t}
dk_x,
\end{equation}
where $\Phi$ is expressed as

\begin{equation}
\Phi = \xi Bk_x^2+(V_x-c_x)k_x,
\quad
V_x=\frac{x}{t}.
\end{equation}
The integral~\ref{eq:app:KP:SampleIntegral} is evaluated through complex contour integration, and contours are chosen such that they move through saddle points; this enables the method of steepest descent to be applied as $t$ approaches infinity.  To apply the method, saddle points, $k_s$, are determined by writing the derivatives of $\Phi$ as

\begin{equation}
\frac{d\Phi}{dk_x}(k_s) = 2\xi Bk_s+(V_x-c_x) = 0,
\end{equation}

\begin{equation}
\frac{d^2\Phi}{dk_x^2}(k_s) = 2\xi B \neq 0,
\end{equation}
to yield

\begin{equation}
\label{eq:app:KP:SaddlePoint}
k_s=\frac{(c_x-V_x)}{2\xi B},
\qquad V_x= \frac{x}{t}.
\end{equation}
In taking the derivative, we hold $V_x$, $B$, and $\xi$ fixed. Note that for any velocity, $\hat{V}$, the saddle point is purely real. The complex integration contour for $\mathcal{Q}_x$ through the saddle point $k_s$ is shown in Fig.~\ref{fig:app:KP:SampleContour}. The angles, $\gamma_1$ and $\gamma_2$ are chosen such that the contours can be evaluated (see Supplemental Material Section~\ref{sup:KP:QWIntegrals}).
\FloatBarrier	
\noindent	
Because there are no poles enclosed in either side of the contour, the following equations must be true:

\begin{equation}
0=~
\int\limits_{C_{1a}}e^{i\Phi t}dk_x
+\int\limits_{C_3b}e^{i\Phi t}dk_x
+\int\limits_{C_4}e^{i\Phi t}dk_x,
\end{equation}

\begin{equation}
0=~
\int\limits_{C_{1b}}e^{i\Phi t}dk_x
+\int\limits_{C_2}e^{i\Phi t}dk_x
+\int\limits_{C_3a}e^{i\Phi t}dk_x.
\end{equation}	
Additionally, the contours are established such that

\begin{equation}
\mathcal{Q}_x=
\lim_{R\to\infty}\left[~
\int\limits_{C_{1a}}e^{i\Phi t}dk_x
+\int\limits_{C_{1b}}e^{i\Phi t}dk_x
~\right].
\end{equation}
Therefore, $\mathcal{Q}_x$ can be evaluated as

\begin{equation}
\label{eq:app:KP:ContoursSolution}
\mathcal{Q}_x=
\lim_{R\to\infty}\left[
-\int\limits_{C_2}e^{i\Phi t}dk_x
-\int\limits_{C_3a}e^{i\Phi t}dk_x
-\int\limits_{C_3b}e^{i\Phi t}dk_x
-\int\limits_{C_4}e^{i\Phi t}dk_x
\right].
\end{equation}

Although motivated by the method of steepest descent to extract long-time behavior, the contours in Fig.~\ref{fig:app:KP:SampleContour} enable the exact evaluation of the integrals in Eq.~(\ref{eq:app:KP:ContoursSolution}) for all time.(see Supplemental Material section \ref{sup:KP:QxIntegral}).
After evaluating Eq.~(\ref{eq:app:KP:ContoursSolution}) and applying the same methodology to $\mathcal{Q}_z$, $\mathcal{W}_x$, and $\mathcal{W}_z$ in Eq.~(\ref{eq:app:KP:QWForms}) above we obtain
\begin{equation}
\mathcal{Q}=
\frac{i\pi}{\xi Bt}
e^{\big(\frac{-i}{4\xi B}\left(
(V_x-c_x)^2
+(V_z-c_z)^2
\right) t\big)
},
\end{equation}	
		
\begin{equation}
\mathcal{W} = 
\frac{-i\pi}{\xi B t}
e^{\big(\frac{it}{4\xi B}
\left(
(V_x-c_x)^2
+(V_z-c_z)^2
\right)\big)}.
\end{equation}

	\subsection{Difference in response breadth between 1D-KRK and 2D-KRK responses}
		\label{app:KP:PeakWidth}		

As mentioned in Section~\ref{sec:KP:DimensionalComparison}, the difference in width between the 1D-KRK and 2D-KRK peaks can be extracted by plotting the data normalized by peak height.
	
\begin{figure}[ht!]
\centering
\includegraphics[keepaspectratio,width=7in]{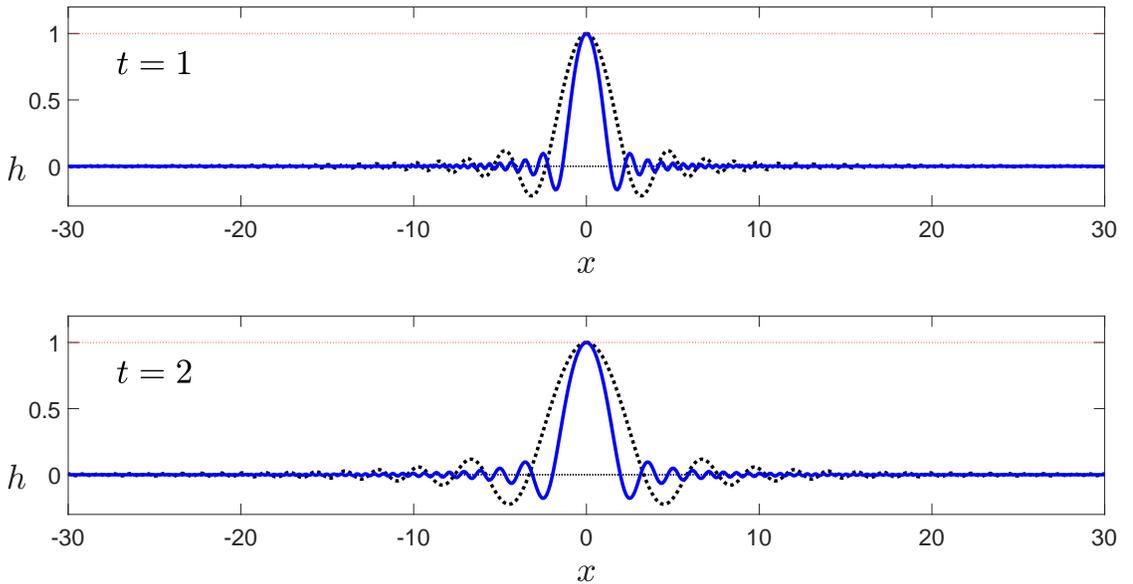}
\caption{A comparison between the 1D solution --Eq.~(\ref{eq:KP:1DSolution})-- (dotted black lines) and the 2D asymptotic solution --Eq.~(\ref{eq:KP:Solution})-- (Solid blue lines) with data normalized to the respective peaks.
Just as with Fig.~\ref{fig:KP:Dimensional_comparison_no_convection}, the data were generated with $A=10$ and $B=1$.}
\label{fig:app:KP:NormalizedPeaks}
\end{figure}
\FloatBarrier			

\noindent
Note that the 1D solution in Fig.~\ref{fig:app:KP:NormalizedPeaks} has nontrivial features that extend farther in the $x$-direction (i.e., has a larger breadth) than that of the 2D solution, as described and justified in Section~\ref{sec:KP:DimensionalComparison}. This difference is further demonstrated in the regions away from $\nicefrac{x}{t}=0$ by normalizing both data sets. The finer structure of the 2D-KRK solution in Fig.~\ref{fig:KP:Dimensional_comparison_no_convection}    are lost because of the the difference in amplitudes between the 1D and 2D solutions.

\section{2D-CMH: Analysis}
	\label{app:NP:CMH}
		
\subsection{Fourier Series Solution (FSS)}
	\label{app:NP:FSS}
	
The FSS for Eq.~(\ref{eq:NP:Operator}) is found using standard methods~\cite{barlow2010} subject to the following initial conditions:	
		

\begin{equation}
h(x,z,0) = 
H_0 \delta(x)\delta(z)
,\qquad
\frac{\partial h}{\partial t}(x,z,0) =
U_0 \delta(x)\delta(z).
\end{equation}
Over the finite domains of  $x\in[-L_x,L_x]$ and $z\in[-L_z,L_z]$, the FSS is

\begin{subequations}
\begin{equation}
FSS(x,z,t) = 
\sum_{j=-\infty}^{\infty}
\sum_{g=-\infty}^{\infty}
f_{j,g}(t)e^{ik_jx}e^{ik_gz},
\qquad k_j = \frac{j\pi}{L_x},
\quad k_g = \frac{g\pi}{L_z},
\end{equation}

\begin{equation}
f_{j,g} = 
e^{-\frac{\eta}{2}t}
\left(
	\frac{1}{4L_xL_z}
\right)
\left(
H_0\cos\left(
\frac{\eta\sqrt{3}}{2}t
\right)
+
2\frac{\psi}{\eta}
\sin\left(
	\frac{\eta\sqrt{3}}{2}t
\right)
\right),
\end{equation}

\begin{equation}
\psi = \frac{1}{\sqrt{3}}
\left(
(A+U_0)+\frac{1}{2}\eta H_0)
\right)
,\qquad
\eta = (k_j^2+k_g^2).
\end{equation}		
\end{subequations}

	\subsection{Long-time asymptotic behaviour of $\theta$ integral in Eq.~(\ref{eq:NP:PolarForm}) }
		\label{app:NP:ThetaIntegral}
		\subsubsection{Phase Function and Saddle Points for Method Of Steepest Descent}
			\label{app:NP:PhaseFunction}
	The integral in $\theta$ from Eq.~\ref{eq:NP:PolarForm} is evaluated asymptotically using the method of steepest descent as $t\to\infty$. To do so, we first rewrite the integral in terms of a phase function $\Phi$, so

\begin{equation}
\label{eq:app:NP:ThetaIntegral}
\mathcal{I}_\theta = 
\int\limits_{0}^{2\pi}
e^{i \Phi t}
d\theta,
\quad
\Phi = \left(
\cos(\theta)V_x
+\sin(\theta)V_z
\right)\xi.
\end{equation}
According to the method, derivatives of $\Phi(\theta)$ are taken to find relevant saddle points.

\begin{equation}
\frac{d\Phi}{d\theta} = \left(
-\sin(\theta)V_x
+\cos(\theta)V_z
\right)\xi,
\end{equation}

\begin{equation}
\frac{d^2\Phi}{d\theta^2} = -\left(
\cos(\theta)V_x
+\sin(\theta)V_z
\right)\xi,
\end{equation}
We observe that $\Phi(\theta)$ has relevant $2^\text{nd}$ order saddle points, $\theta_s$,  where the first derivative is zero and the second derivative is non-zero. These points are defined as

\begin{equation}
\label{eq:app:NP:ThetaSaddles}
\theta_s = \arctan
\left(
\frac{V_z}{V_x}
\right)+n\pi.
\end{equation}
Note that, for $m=1,2,3,...$, the odd derivatives, $2m-1$, are zero at $\theta=\theta_s$ and the signs of the even derivatives alternate as $(-1)^m \left(\cos(\theta_s)V_x + \sin(\theta_s)V_z\right)$.
According to the method of steepest descent, the asymptotic behavior of the integral --Eq.~(\ref{eq:app:NP:ThetaIntegral})-- will be dominated by the region near the saddle point.  To this end, the phase function $\Phi$ is linearized near the saddle point as

\begin{equation}
\label{eq:app:NP:LinearizedPhiIntermediate}
\Phi_{\theta_s} =
\Phi(\theta_s)
\left(
\sum_{n=0}^{\infty}
\frac{(-1)^n}{(2n)!}
(\theta-\theta_s)^{2n}
\right) .
\end{equation}
In Eq.~(\ref{eq:app:NP:LinearizedPhiIntermediate}), note that $\Phi_{\theta_s}$ denotes the Taylor series expansion of $\Phi$ near the saddle point, and $\Phi^{(n)}(\theta_s)$ denotes the $n^\text{th}$ derivative of $\Phi$ evaluated at the saddle point. Dropping the higher orders of $(\theta-\theta_s)$, we are left with

\begin{equation}
\label{eq:app:NP:LinearizedPhi}
\Phi_{\theta_s} \sim 
\Phi(\theta_s)
\left(
1-\frac{1}{2}
(\theta-\theta_s)^2
+O\big[(\theta-\theta_s)^4\big]
\right)
\quad \text{ as }
\theta\to\theta_s,
\end{equation}
which is sufficient to extract the leading order behaviour of the integral --Eq.~(\ref{eq:app:NP:ThetaIntegral})-- in what follows.

		\subsubsection{Integration contours for $\theta$ integral in solution of 2D-CMH problem}
			\label{app:NP:Contours}
The following is typical of the contours which are established to evaluate Eq.~(\ref{eq:app:NP:ThetaIntegral}) and the analysis thereof. The specific layout of the phase space depends on the signs and values of $V_x$ and $V_z$. The following analysis assumes that the saddle points, $\theta_s$ do not lie on the ends of the integration contour. The case where they do is provided in Supplemental Material section \ref{sup:NP:SpecialCaseTheta}. 

\begin{figure}[ht!]
\centering
\includegraphics[keepaspectratio,width=6in]{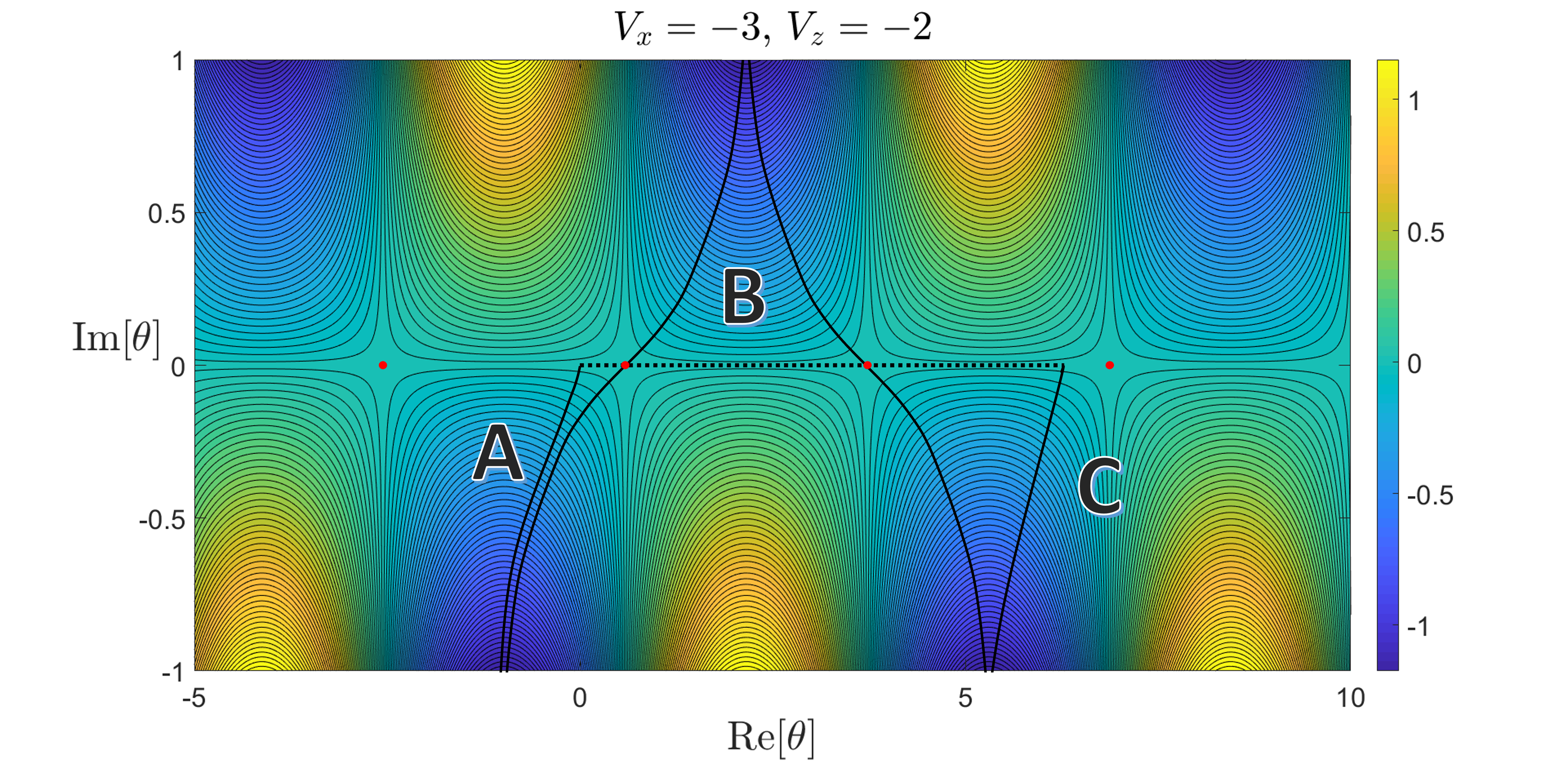}
\caption{Contours of constant Re$[i\Phi]$ as a function of Re$[\theta]$ and Im$[\theta]$ from Eq.~(\ref{eq:app:NP:ThetaIntegral}). The integral of interest is marked with a dotted line running between $\theta = 0$ and $\theta = 2\pi$. Saddle points at $\theta_s=\arctan\left(\nicefrac{V_z}{V_x}\right)+n\pi$ are marked with (\textcolor{red}{$\bullet$}). Three closed contours are established which each contain a portion of the integral of interest. Each saddle point is at the maximum value of the contour, with each contour closing as Im$[\theta]\to\pm\infty$ which equates to Re$[i\Phi]\to -\infty$. Each contour will be generalized as three triangles in Fig.~\ref{fig:app:NP:Triangles}.}
\label{fig:app:NP:SaddleMap}
\end{figure}
\FloatBarrier
\noindent
For the following analysis, the contours are generalized to be three triangles as shown below.
	
\begin{figure}[ht!]
\centering
\includegraphics[keepaspectratio,width=5in]{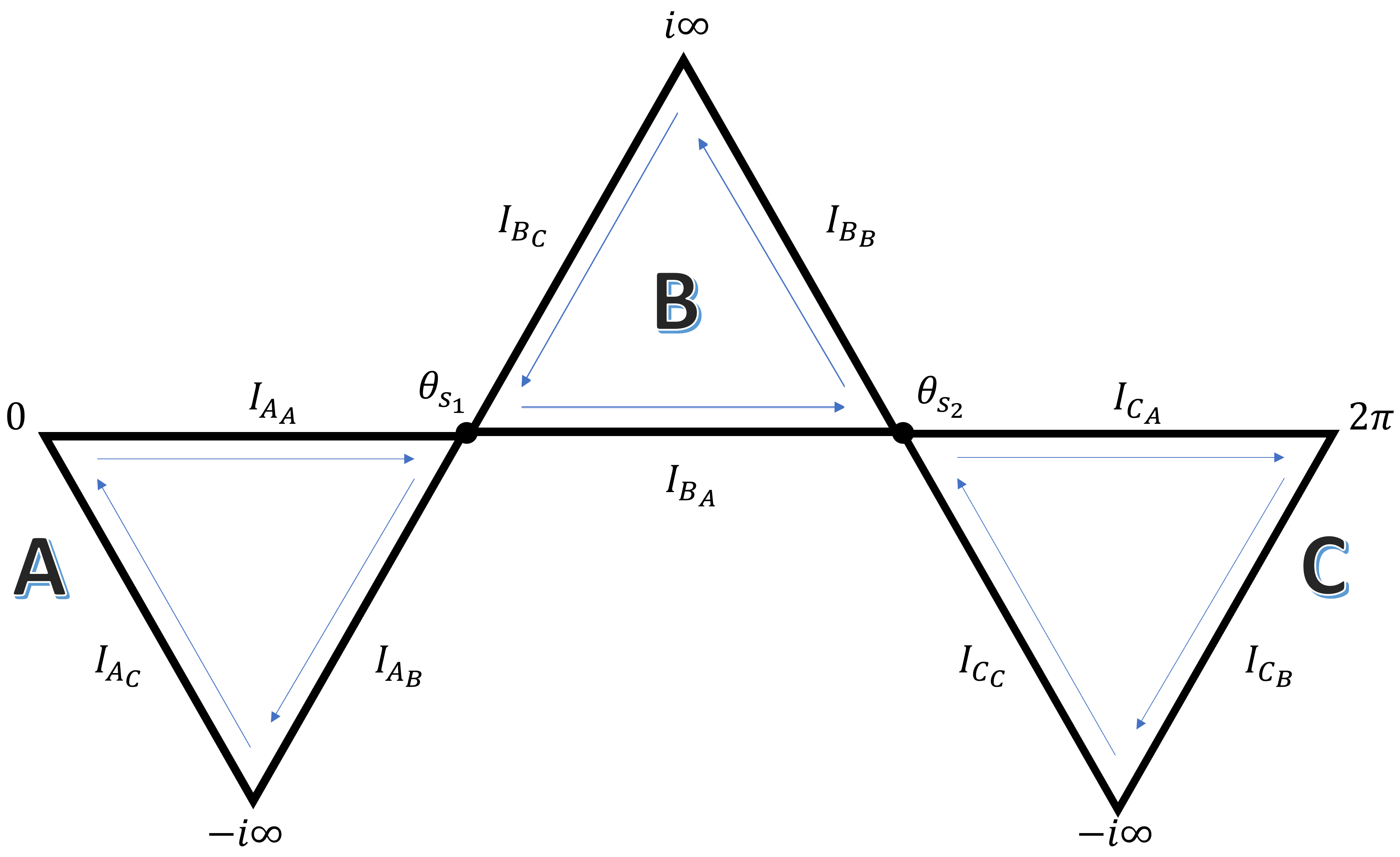}
\caption{Simplified schematic of the contours in Fig.~\ref{fig:app:NP:SaddleMap}. The integration contour for Eq.~(\ref{eq:app:NP:ThetaIntegral}) is given by the sub-contours $I_{A_A}$, $I_{B_A}$, and $I_{C_A}$. The remaining sub-contours serve to form three closed contours, $A$, $B$, and $C$. The points $\theta_{S_1}$ and $\theta_{S_2}$ are two saddle points which lie on the integration contour. Note that two saddles will always lie on the $\theta$ contour if $\nicefrac{V_z}{V_x}\neq 0$ as per Eq.~(\ref{eq:app:NP:ThetaSaddles}).
}
\label{fig:app:NP:Triangles}
\end{figure}
\FloatBarrier

Because there are no poles (enclosed or otherwise), the closed contours in Figs.~\ref{fig:app:NP:Triangles} lead to the following equations:

\begin{subequations}
\begin{equation}
0=\oint_A=\mathcal{I}_{A_A}+\mathcal{I}_{A_B}+\mathcal{I}_{A_C},
\end{equation}

\begin{equation}
0=\oint_B=\mathcal{I}_{B_A}+\mathcal{I}_{B_B}+\mathcal{I}_{B_C},
\end{equation}

\begin{equation}
0=\oint_C=\mathcal{I}_{C_A}+\mathcal{I}_{C_B}+\mathcal{I}_{C_C},
\end{equation}
\end{subequations}
which can be rearranged as 

\begin{subequations}
\begin{equation}
\mathcal{I}_{A_A}=
-\mathcal{I}_{A_B}
-\mathcal{I}_{A_C},
\end{equation}

\begin{equation}
\mathcal{I}_{B_A}=
-\mathcal{I}_{B_B}
-\mathcal{I}_{B_C},
\end{equation}

\begin{equation}
\mathcal{I}_{C_A}=
-\mathcal{I}_{C_B}
-\mathcal{I}_{C_C}.
\end{equation}
\end{subequations}
As we are considering the cases where a saddle point does not lie at $0$ or $2\pi$ (the limits of integration for Eq.~(\ref{eq:app:NP:ThetaIntegral})), it is observed that contours without saddle points are asymptotically subdominant to those with saddle points as $t$ goes off to infinity. The equations can be written as

\begin{subequations}
\begin{equation}
\mathcal{I}_{A_A}\sim
-\mathcal{I}_{A_B}
\quad
\text{ as }
t\to\infty,
\end{equation}

\begin{equation}
\mathcal{I}_{B_A}=
-\mathcal{I}_{B_B}
-\mathcal{I}_{B_C},
\end{equation}

\begin{equation}
\mathcal{I}_{C_A}\sim
-\mathcal{I}_{C_C}
\quad
\text{ as }
t\to\infty.
\end{equation}
\end{subequations}		
The total integral in $\theta$ can therefore be written asymptotically as 

\begin{equation}
\label{eq:app:NP:ThetaAsympFormGeneral}
\mathcal{I}_\theta\sim
-\mathcal{I}_{A_B}
-\mathcal{I}_{B_B}
-\mathcal{I}_{B_C}
-\mathcal{I}_{C_C}
\quad
\text{ as }
t\to\infty.
\end{equation}	
The integral along a sub-contour ($I_{A_B}$, $I_{B_C}$, $I_{B_B}$, or $I_{C_C}$), denoted here as $\int_S$, can be written in  terms of the radial distance $r$ and angle $\gamma$ relative to a saddle point. This is done by letting $\theta = \theta_s + r e^{i\gamma}$, $d\theta = e^{i\gamma}dr$.
Fundamentally, each sub-integral can be broken up into two regions, the region near the saddle point, and the rest of the path.

Along a sub-contour, the region near the saddle point is asymptotically dominant over the rest of the path though integration by parts. Additionally, when looking near the saddle point, $\Phi$ can be linearized using Eq.~(\ref{eq:app:NP:LinearizedPhi}).
Then, the upper bound of integration can be increased to infinity, accruing only asymptotically subdomninant terms~\cite{bender1999}. In the conversion to radial coordinates, the angle $\gamma$ is the angle of approach or departure of the steepest contours through the saddle point; the steepest paths are perpendicular to the contours of constant Re$[i\Phi]$ in the complex $\theta$ plane.
The values of $\gamma$ are chosen such that the integrand takes the form of $e^{-r^2}dr$.
Using ``$\pm$" to denote that the sign depends on the particular sub-contour being examined,
\begin{equation}
\int\limits_{S} \sim
\pm e^{i\Phi(\theta_s)t+i\gamma}
\int_0^\infty
e^{-\frac{\Phi(\theta_s)}{2} r^2 t}dr
\quad
\text{ as }
t\to\infty,
\end{equation}
which evaluates to

\begin{equation}
\int\limits_{S} \sim
\pm e^{i\Phi(\theta_s)t+i\gamma}
\sqrt{\frac{\pi}{2\Phi(\theta_s) t}}
\quad
\text{ as }
t\to\infty.
\end{equation}
Applying the method to each sub-contour in Eq.~(\ref{eq:app:NP:ThetaAsympFormGeneral}), the overall integral in $\theta$ is expressed asymptotically as
		
\begin{equation}
\mathcal{I}_\theta \sim
2\cos
\left(
\xi\hat{V}t-\frac{\pi}{4}
\right)
\sqrt{\frac{2\pi}{\hat{V} t}}
\quad \text{ as } t\to\infty,
\end{equation}		
where

\begin{equation}
\hat{V}=\sqrt{V_x^2+V_z^2}.
\end{equation}
Note that when the saddle points lie on the ends of the integration contour, the solution is equivalent (Supplemental Material Section \ref{sup:NP:SpecialCaseTheta}).

		
	\subsection{Criterion for agreement between asymptotic and FSS solution}
		\label{app:NP:DivergenceCriterion}
\begin{figure}[ht!]
\centering
\includegraphics[keepaspectratio,width=7in]{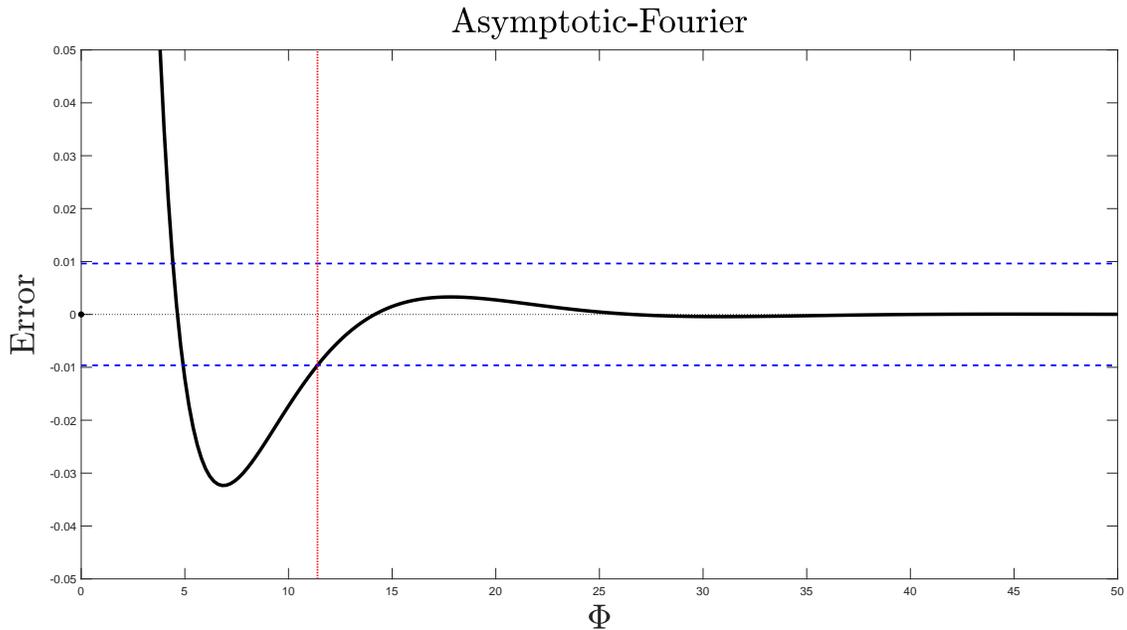}
\caption{The difference between the asymptotic solution --Eq.~(\ref{eq:NP:AsymptoticSolFull})-- and the FSS (\ref{app:NP:FSS}) versus $\Phi$. The blue dashed lines denote where the difference between the two solutions is equal to $\pm 1\%$ of the exact peak height in Eq.~(\ref{eq:NP:V0SolutionPhi}). A vertical red line marks the value of $\Phi=11.4$ above which the error is between the marked bounds. The black dotted line marks the $\Phi$ axis.
}
\label{fig:app:NP:DivergenceCriterion}
\end{figure}

As mentioned at the end of Section~\ref{sec:NP:CMH}, the judgment of when $\Phi$ is sufficiently large for the FSS~(\ref{app:NP:FSS}) and asymptotic solution --Eq.~(\ref{eq:NP:AsymptoticSolFull})-- to agree is made based on Fig.~\ref{fig:app:NP:DivergenceCriterion}.
Here, one can observe that for $\Phi\geq 11.4$, indicated in Fig.~\ref{fig:app:NP:DivergenceCriterion}, the magnitude of the error is less than $\nicefrac{A}{(600\sqrt{3})}\approx 0.01$ ($1\%$ of the peak height given by Eq.~(\ref{eq:NP:V0SolutionPhi})).
\FloatBarrier

\end{appendix}

\newpage

\section{ Supplemental Material: 2D-KRK: Analysis}
	\label{sup:KP:KRK}
	
	\subsection{Evaluating $h$ for initial height disturbance}
		\label{sup:KP:InitialConditions}
				
Eq.~(\ref{eq:app:KP:H0Integral}) is transcribed below for completeness.		
		
		\begin{equation*}
h = H_0
\int\limits_{-\infty}^{\infty}
\int\limits_{-\infty}^{\infty}
\left(
\left(
\frac{\psi + \eta}
{\eta}
\right)
\left( 
\frac{
	e^{i\eta t}
	-
	e^{-i\eta t}
	}{2}
\right)
+e^{-i\eta t}
\right)
e^{-i\psi t}
e^{ik_xV_xt}
e^{ik_zV_zt}
dk_zdk_x,
\end{equation*}

\begin{equation}
\label{eq:sup:KP:H0Integral}
\psi = c_xk_x + c_zk_z,
\quad
\eta = k_x^2+k_z^2,
\quad
V_x = \frac{x}{t},
\quad
V_z = \frac{z}{t}.
\end{equation}
Eq.~(\ref{eq:sup:KP:H0Integral}) is broken up as

\begin{equation}
\label{eq:sup:KP:ABCForm}
h(x,z,t)=\frac{H_0}{4\pi^2}
(\mathcal{A}+\mathcal{B}+\mathcal{C}),
\end{equation}

\begin{equation}
\mathcal{A}=
\int\limits_{-\infty}^{\infty}
\int\limits_{-\infty}^{\infty}
e^{-i\psi t}
\Bigg(
\frac{i\psi(e^{i\eta t}-e^{- i\eta t})}{2i\eta}
\Bigg)
e^{ik_xx}
e^{ik_zz}
dk_zdk_x,
\end{equation}

\begin{equation}
\label{eq:sup:KP:BIntegral}
\mathcal{B}=
\int\limits_{-\infty}^{\infty}
\int\limits_{-\infty}^{\infty}
e^{-i\psi t}
\Bigg(
\frac{i\eta(e^{i\eta t}-e^{- i\eta t})}{2i\eta}
\Bigg)
e^{ik_xx}
e^{ik_zz}
dk_zdk_x,
\end{equation}

\begin{equation}
\mathcal{C}=
\int\limits_{-\infty}^{\infty}
\int\limits_{-\infty}^{\infty}
e^{-i\psi t}
\Bigg(
 e^{- i\eta t}
\Bigg)
e^{ik_xx}
e^{ik_zz}
dk_zdk_x,
\end{equation}

\begin{equation}
\psi = c_xk_x + c_zk_z,
\quad
\eta = k_x^2+k_z^2.
\end{equation}

\subsubsection{Solution to $\mathcal{A}$ (sub-integral of $h$)}
	\label{sup:KP:SubIntegralA}
The integral $\mathcal{I}$ is established such that 

\begin{equation}
\label{eq:sup:KP:IIntegral}
\mathcal{I}=\frac{-1}{t}
\int\limits_{-\infty}^{\infty}
\int\limits_{-\infty}^{\infty}
e^{-i \gamma \psi t}
\Bigg(
\frac{\sin(\eta t)}{\eta}
\Bigg)
e^{ik_xx}
e^{ik_zz}
dk_zdk_x,
\end{equation}

\begin{equation}
\label{eq:sup:KP:RelateIAndA}
\frac{\partial\mathcal{I}}{\partial\gamma}\Big|_{\gamma=1}=\mathcal{A}.
\end{equation}
\noindent
Another parameter, $\alpha$, is introduced into Eq.~(\ref{eq:sup:KP:IIntegral}) such that

\begin{equation}
\hat{\mathcal{I}} = 
\frac{-1}{t}
\int\limits_{-\infty}^{\infty}
\int\limits_{-\infty}^{\infty}
e^{-i \gamma \psi t}
\Bigg(
\frac{\sin(\alpha\eta t)}{\eta}
\Bigg)
e^{ik_xx}
e^{ik_zz}
dk_zdk_x,
\end{equation}

\begin{equation}
\hat{\mathcal{I}}\big|_{\alpha = 0} = 0,
\quad
\hat{\mathcal{I}}\big|_{\alpha = 1} = \mathcal{I}.
\end{equation}
The derivative is taken with respect to $\alpha$,

\begin{equation}
\frac{\partial\hat{\mathcal{I}}}{\partial\alpha} = 
-\int\limits_{-\infty}^{\infty}
\int\limits_{-\infty}^{\infty}
e^{-i \gamma \psi t}
\Bigg(
\cos(\alpha\eta t)
\Bigg)
e^{ik_xx}
e^{ik_zz}
dk_zdk_x.
\end{equation}
Through the same steps as laid out for $\nicefrac{\partial h}{\partial \xi}$ in Section~\ref{sec:KP:IntegralSolution}, we find that

\begin{equation*}
\frac{\partial\hat{\mathcal{I}}}{\partial\alpha} = \frac{-\pi}{\alpha B t}
\sin \left(
\frac{t}{4\alpha B}
\left(
\hat{V}_{\gamma}
\right)
\right),
\end{equation*}

\begin{equation}
\hat{V}_{\gamma} = (V_x-\gamma c_x)^2+(V_z-\gamma c_z)^2.
\end{equation}
Integrating in $\alpha$ yields

\begin{equation}
\mathcal{I} = 
\frac{\pi}{Bt}
\left(
\int\limits_{0}^{\frac{\bar{V}_{\gamma}t}{4B}}
\frac{1}{u}
\sin(u)dv
-
\frac{\pi}{2}
\right),
\end{equation}
and differentiating in $\gamma$ yields,

\begin{equation}
\frac{\partial\mathcal{I}}{\partial\gamma} = 
-\Big(
c_x(V_x-\gamma c_x)+c_z(V_z-\gamma c_z)
\Big)
\frac{2\pi}{\bar{V}_{\gamma}Bt}
\sin
\left(\frac{\bar{V}_{\gamma}t}{4B}\right).
\end{equation}
From the construction of $\mathcal{I}$ in Eq.~(\ref{eq:sup:KP:RelateIAndA}),

\begin{equation}
\mathcal{A}=
-\Big(
c_x(V_x- c_x)+c_z(V_z- c_z)
\Big)
\frac{2\pi}{\bar{V}Bt}
\sin
\left(\frac{\bar{V}t}{4B}\right)
\end{equation}

\subsubsection{Solution to $\mathcal{B}$ (sub-integral of $h$)}
	\label{sup:KP:SubIntegralB}
Eq.~(\ref{eq:sup:KP:BIntegral}) is rearranged into the form
\begin{multline}
\mathcal{B}=
\frac{1}{2}
\int\limits_{-\infty}^{\infty}
e^{(iBk_x^2+i(V_x-c_x)k_x)t}
\int\limits_{-\infty}^{\infty}
e^{(iBk_z^2+i(V_z-c_z)k_z)t}
dk_zdk_x
\\-
\frac{1}{2}
\int\limits_{-\infty}^{\infty}
e^{(-iBk_x^2+i(V_x-c_x)k_x) t}
\int\limits_{-\infty}^{\infty}
e^{(-iBk_z^2+i(V_z-c_z)k_z) t}
dk_zdk_x
\end{multline}
Combined with Eqs.~(\ref{eq:KP:QIntegral}) and~(\ref{eq:KP:WIntegral}, it is shown in Section~\ref{sup:KP:QWIntegrals} that

\begin{equation}
\mathcal{Q}=
\int\limits_{-\infty}^{\infty}
e^{\left(i\xi Bk_z^2+i(\frac{z}{t}-c_z)k_z \right)t}
\left[
\int\limits_{-\infty}^{\infty}
e^{\left(i\xi Bk_x^2+i(\frac{x}{t}-c_x)k_x \right)t}
dk_x
\right]dk_z=
\frac{i\pi}{\xi Bt}
e^{\big(\frac{-it}{4\xi B}\left(
(c_x-V_x)^2
+(c_z-V_z)^2
\right)\big)},
\end{equation}

\begin{equation}
\mathcal{W}=
\int\limits_{-\infty}^{\infty}
e^{\left(-i\xi Bk_z^2+i(\frac{z}{t}-c_z)k_z\right)t}
\left[
\int\limits_{-\infty}^{\infty}
e^{\left(-i\xi Bk_x^2+i(\frac{x}{t}-c_x)k_x \right)t}
dk_x
\right]dk_z= 
\frac{-i\pi}{\xi B t}
e^{\big(\frac{it}{4\xi B}
\left(
(V_x-c_x)^2
+(V_z-c_z)^2
\right)\big)}.
\end{equation}
Therefore,

\begin{equation}
\mathcal{B}=
\frac{i\pi}{Bt}
\cos\left(
\frac{t}{4B}
((V_x-c_x)^2+(V_z-c_z)^2)\right).
\end{equation}

\subsubsection{Solution to $\mathcal{C}$ (sub-integral of $h$)}
	\label{sup:KP:SubIntegralC}
Using results in Section~\ref{sup:KP:QWIntegrals} to follow and combining with Eq.~(\ref{eq:KP:QIntegral}), we obtain

\begin{multline}
\mathcal{W}=
\int\limits_{-\infty}^{\infty}
e^{\left(-i\xi Bk_z^2+i(\frac{z}{t}-c_z)k_z\right)t}dk_z
\int\limits_{-\infty}^{\infty}
e^{\left(-i\xi Bk_x^2+i(\frac{x}{t}-c_x)k_x \right)t}
dk_x
\\= 
\frac{-i\pi}{\xi B t}
e^{\big(\frac{it}{4\xi B}
\left(
(V_x-c_x)^2
+(V_z-c_z)^2
\right)\big)},
\end{multline}

\begin{equation}
\mathcal{C}= 
\frac{-i\pi}{ B t}
e^{\big(\frac{it}{4 B}
\left(
(V_x-c_x)^2
+(V_z-c_z)^2
\right)\big)}.
\end{equation}

\subsubsection{Assembling sub-integrals to obtain the solution to $h$}
	Substituting the results from Sections~\ref{sup:KP:SubIntegralA}, \ref{sup:KP:SubIntegralB}, and \ref{sup:KP:SubIntegralC} into Eq.~\ref{eq:sup:KP:ABCForm} yields
	
\begin{multline}
h(x,z,t)=\frac{H_0}{4\pi^2}
\Big(
-\Big(
	c_x(V_x- c_x)+c_z(V_z- c_z)
	\Big)
	\frac{2\pi}{\bar{V}Bt}
	\sin
	\left(\frac{\bar{V}t}{4B}\right)
\\+\frac{i\pi}{Bt}
	\cos\left(
	\frac{t}{4B}
	((V_x-c_x)^2+(V_z-c_z)^2)\right)
+\frac{-i\pi}{ B t}
	e^{\big(\frac{it}{4 B}
	\left(
		(V_x-c_x)^2
		+(V_z-c_z)^2
	\right)\big)}
 \Big),
\end{multline}
which can be rearranged into the form found in Eq.~(\ref{eq:app:KP:H0Sol}) in the appendix, transcribed below for completeness

\begin{equation*}
h(x,z,t)=
\frac{H_0}{4Bt\pi}
\Bigg(
1
-2
\frac{c_x(V_x-c_x)+c_z(V_z-c_z)}{\hat{V}}
\Bigg)
\sin
\left(\frac{\hat{V}t}{4B}\right),
\end{equation*}

\begin{equation}
\hat{V} = \sqrt{(V_x-c_x)^2+(V_z-c_z)^2}.
\end{equation}

	\subsection{Development of Fourier inversion integral for 2D-KRK operator}
		\label{sup:KP:FourierIntegral}

Taking the Fourier transforms in $x$ and $z$ of the 2D-KRK operator --Eq.~(\ref{eq:KP:Operator})-- yields a second order ordinary differential equation of the Fourier transformed variable, $\hat{\hat{h}}_{xz}(t)$.

\begin{equation}
\label{eq:sup:KP:FourierODE}
\hat{\hat{h}}_{xz}=
\frac{A}{\eta}e^{-i\psi t}\left(
\frac{e^{i\eta t}-e^{-i\eta t}}
{2i}
\right)
,\qquad
\psi = c_xk_x + c_zk_z
,\qquad
\eta = B(k_x^2+k_z^2), 
\end{equation}

\begin{equation}
\hat{\hat{h}}_{xz}(0)=0,
\qquad
\frac{d \hat{\hat{h}}_{xz}}{dt}(0)=0,
\end{equation}
	
\begin{equation}
\psi = c_xk_x + c_zk_z,
\qquad
\eta = B(k_x^2+k_z^2).
\end{equation}
Note that, in this construction, $\psi$ and $\eta$ are always real and $\eta$ is non-negative.
The inverse Fourier transforms (in $x$ and $z$) are reproduced here from the main text Eq.~(\ref{eq:KP:FourierSolution}) for completeness.

\begin{equation}
h(x,z,t) = 
\frac{1}{4\pi^2}
\int\limits_{-\infty}^{\infty}
\int\limits_{-\infty}^{\infty}
\hat{\hat{h}}_{xz}
e^{ik_xx}
e^{ik_zz}
dk_zdk_x,
\end{equation}


\begin{equation}
h(x,z,t) = 
\frac{A}{4\pi^2}
\int\limits_{-\infty}^{\infty}
\int\limits_{-\infty}^{\infty}
e^{-i\psi t}\left(
\frac{\sin(\eta t)}
{\eta}
\right)
e^{ik_xx}
e^{ik_zz}
dk_zdk_x
\qquad \eta = B(k_x^2 + k_z^2).
\end{equation}

	\subsection{Evaluating inversion integral through introduction of parameter $\xi$}
		\label{sup:KP:IntroduceXi}

The variable $\xi$ is introduced to create the function $\tilde{h}(x,z,t,\xi)$ such that

\begin{equation}
\label{eq:sup:KP:hTildeConditions}
\tilde{h}(x,z,t,0) = 0,
\qquad
\tilde{h}(x,z,t,1) = h(x,z,t),
\end{equation}

\begin{equation}
\tilde{h}(x,z,t,\xi) = 
\frac{A}{4\pi^2}
\int\limits_{-\infty}^{\infty}
\int\limits_{-\infty}^{\infty}
e^{-i\psi t}\left(
\frac{\sin(\xi \eta t)}
{\eta}
\right)
e^{ik_xx}
e^{ik_zz}
dk_zdk_x.
\end{equation}
The derivative is taken with respect to $\xi$,

\begin{equation}
\label{eq:sup:KP:dhdxiForm}
\frac{\partial\tilde{h}}{\partial \xi} = 
\frac{At}{4\pi^2}
\int\limits_{-\infty}^{\infty}
\int\limits_{-\infty}^{\infty}
e^{-i\psi t}\cos(\xi B(k_x^2+k_z^2)t)
e^{ik_xx}
e^{ik_zz}
dk_zdk_x,
\end{equation}
and thus, in accordance with Eq.~(\ref{eq:sup:KP:hTildeConditions}), we have

\begin{equation}
\label{eq:sup:KP:XiIntegral}
h(x,z,t) = \int\limits_0^1 \frac{\partial\tilde{h}}{\partial\xi}d\xi.
\end{equation}	
The function $\nicefrac{\partial\tilde{h}}{\partial\xi}$ is evaluated as laid out in~\ref{app:KP:ContourIntegration} resulting in the following:

\begin{equation}
\label{eq:sup:KP:dhdxiSolution}
\frac{\partial\tilde{h}}{\partial \xi}=
\frac{A }{4\xi B\pi }
\sin\left(
\frac{t}{4\xi B}\left(
(V_x-c_x)^2
+(V_z-c_z)^2
\right)
\right).
\end{equation}	
Through Eq.~(\ref{eq:sup:KP:XiIntegral}), we obtain

\begin{equation}
h(x,z,t)=
\int\limits_0^1
\frac{A }{4\xi B\pi }
\sin\left(
\frac{t}{4\xi B}\left(
(V_x-c_x)^2
+(V_z-c_z)^2
\right)
\right)
d\xi,
\end{equation}
which can be rearranged to

\begin{equation*}
h(x,z,t)=
\frac{A}{4B\pi}
\left(
\frac{\pi}{2}-
\int\limits_{0}^{\frac{\hat{V}t}{4B}}
\frac{1}{u}
\sin(u)du
\right),
\end{equation*}

\begin{equation}
\label{eq:sup:KP:Solution}
\hat{V}=\sqrt{(V_x-c_x)^2+(V_z-c_z)^2}.
\end{equation}
Note that, when Eq.~(\ref{eq:sup:KP:Solution}) is written in terms of $\Phi=\hat{V}^2t$, the result is Eq.~(\ref{eq:KP:Solution}).

\subsection{Evaluation of $\mathcal{Q}$ and $\mathcal{W}$ integrals}
	\label{sup:KP:QWIntegrals}
	
According to Eq.~(\ref{eq:app:KP:QWForms}), the $\mathcal{Q}$ and $\mathcal{W}$ integrals consist of the product of an integral in $x$ and an integral in $z$. Each sub-integral is handled separately in the following sections.	
	
	\subsubsection{Evaluation of $\mathcal{Q}_x$ sub-integral}
		\label{sup:KP:QxIntegral}
		Working from the analysis in~\ref{app:KP:ContourIntegration}, the integrals in the following expression
\begin{equation}
\mathcal{Q}_x=
\lim_{R\to\infty}\left[
-\int\limits_{C_2}e^{i\Phi t}dk_x
-\int\limits_{C_3a}e^{i\Phi t}dk_x
-\int\limits_{C_4}e^{i\Phi t}dk_x
-\int\limits_{C_3b}e^{i\Phi t}dk_x
\right]
\end{equation}	
need to be evaluated.
Because the angles depicted in Fig.~\ref{fig:app:KP:SampleContour} are limited to be in the intervals

\begin{equation}
\label{eq:sup:KP:GammaValues}
\gamma_1\in
\left[
	0,\frac{\pi}{2}
\right],
\quad
\gamma_2 \in
\left[
	\pi,\frac{3\pi}{2}
\right],
\end{equation}
the integral along contours $C_2$ and $C_4$ go to zero as $R$ goes to infinity as Jordan arcs.
The process for evaluating the two remaining contours is the same for both $C_{3a}$ and $C_{3b}$ (differing only in the direction of integration and value of $\gamma$). Therfore, only the evaluation along $C_{3a}$ will be transcribed below.
The integrand is rotated and shifted through the substitution of $k = k_s + r e^{i\gamma}$, $dk = e^{i\gamma}dr$ to yield

\begin{equation}
\int\limits_{C_3a}e^{i\Phi t}dk_x=
\int\limits_{\infty}^{0}
e^{\left(
-i\frac{(c_x-V_x)^2}{4\xi B}
\right) t}
e^{\left(
i\xi Br^2 e^{2i\gamma_1}
\right) t}
e^{i\gamma_1}dr.
\end{equation}
The value of $\gamma_1$ in Eq.~(\ref{eq:sup:KP:GammaValues}) is chosen to be $\nicefrac{\pi}{4}$, to obtain

\begin{equation}
\int\limits_{C_3a}e^{i\Phi t}dk_x=
-e^{i\frac{\pi}{4}}
e^{\left(
-i\frac{(c_x-V_x)^2}{4\xi B}
\right) t}
\int\limits_{0}^{\infty}
e^{\left(
-\xi Br^2
\right) t}
dr,
\end{equation}

\begin{equation}
\int\limits_{C_3a}e^{i\Phi t}dk_x=
-\frac{1}{2}e^{i\frac{\pi}{4}}
e^{\left(
-i\frac{(c_x-V_x)^2}{4\xi B}
\right) t}
\sqrt{\frac{\pi}{\xi Bt}}.
\end{equation}
Applying the same process to the other contour yields

\begin{equation}
\mathcal{Q}_x=
e^{i\frac{\pi}{4}}
e^{\left(
-i\frac{(c_x-V_x)^2}{4\xi B}
\right) t}
\sqrt{\frac{\pi}{\xi Bt}}.
\end{equation}

	\subsubsection{Evaluation of $\mathcal{Q}_z$ sub-integral}
		The structure of the integral $\mathcal{Q}_z$ is identical to that of $\mathcal{Q}_x$ with $z$ instead of $x$. Therefore,
		
	\begin{equation}
\mathcal{Q}_z=
e^{i\frac{\pi}{4}}
e^{\left(
-i\frac{(c_z-V_z)^2}{4\xi B}
\right) t}
\sqrt{\frac{\pi}{\xi Bt}}.
\end{equation}	
	
	\subsubsection{Evaluation of $\mathcal{W}_x$ sub-integral}
The sub-integral $\mathcal{W}_x$ is	
\begin{equation}
\label{eq:sup:KP:SampleIntegralW}
\mathcal{W}_x=
\int\limits_{-\infty}^{\infty}
e^{\left(-i\xi Bk_x^2+i(\frac{x}{t}-c_z)k_x\right)t}
dk_x.
\end{equation}   	
The integrand of~(\ref{eq:sup:KP:SampleIntegralW}) is expressed in the following form,

\begin{equation}
\mathcal{W}_x=
\int\limits_{-\infty}^{\infty}
e^{i\left(\Phi(k_x) \right)t}
dk_x
\end{equation}
where

\begin{equation}
\Phi = -\xi Bk_x^2+(V_x-c_x)k_x,
\quad
V_x=\frac{x}{t}.
\end{equation}

The derivatives of $\Phi$ are
\begin{equation}
\frac{d\Phi}{dk_x} = -2\xi Bk_x+(V_x-c_x),
\end{equation}

\begin{equation}
\frac{d^2\Phi}{dk_x^2} = -2\xi B.
\end{equation}

\begin{figure}[ht!]
\centering
\includegraphics[keepaspectratio,width=4in]{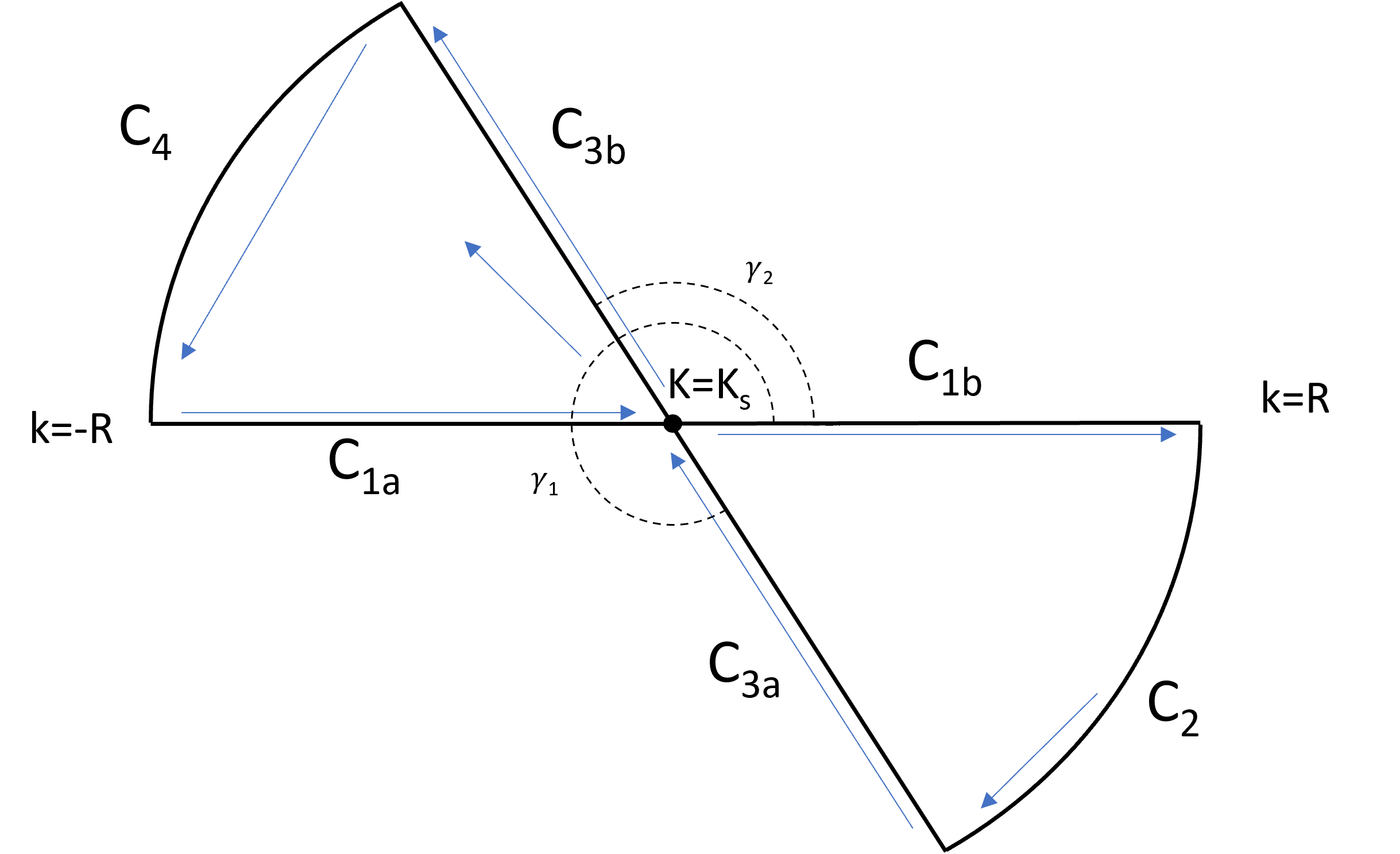}
\caption{Integration path for $\mathcal{W}_x$. Note that, through equation~(\ref{eq:sup:KP:SaddlePointW}) the saddle point $k_s$ always lies on the real axis.}
\label{fig:sup:KP:SampleContourW}
\end{figure}

The function $\Phi$ has a second order saddle point, $k_s$ at

\begin{equation}
\label{eq:sup:KP:SaddlePointW}
k_s=\frac{(V_x-c_x)}{2\xi B},
\qquad V_x= \frac{x}{t}
\end{equation}
Note that, for any velocity ray, the saddle point is purely real. A typical integration contour takes the form of Fig~\ref{fig:sup:KP:SampleContourW} centered on the saddle point $k_s$

\FloatBarrier		
\noindent
Because there are no poles enclosed in either contour, the following must be true from Cauchy's theorem:

\begin{equation}
0=~
\int\limits_{C_{1a}}e^{i\Phi t}dk_x
+\int\limits_{C_3a}e^{i\Phi t}dk_x
+\int\limits_{C_4}e^{i\Phi t}dk_x,
\end{equation}

\begin{equation}
0=~
\int\limits_{C_{1b}}e^{i\Phi t}dk_x
+\int\limits_{C_2}e^{i\Phi t}dk_x
+\int\limits_{C_3b}e^{i\Phi t}dk_x.
\end{equation}
Additionally,

\begin{equation}
\mathcal{W}_x=
\lim_{R\to\infty}\left[
\int\limits_{C_{1a}}e^{i\Phi t}dk_x
+\int\limits_{C_{1b}}e^{i\Phi t}dk_x
\right].
\end{equation}
Therefore, $\mathcal{W}_x$ can be evaluated as

\begin{equation}
\label{eq:sup:KP:ContoursSolutionW}
\mathcal{W}_x=
\lim_{R\to\infty}\left[
-\int\limits_{C_2}e^{i\Phi t}dk_x
-\int\limits_{C_3a}e^{i\Phi t}dk_x
-\int\limits_{C_2}e^{i\Phi t}dk_x
-\int\limits_{C_3b}e^{i\Phi t}dk_x
\right].
\end{equation}
The evaluation of Eq.~(\ref{eq:sup:KP:ContoursSolutionW}) is done with the same process as applied in Section~\ref{sup:KP:QxIntegral}, except that the angles $\gamma_1$ and $\gamma_2$ in Fig.~\ref{fig:sup:KP:SampleContourW} lie in the ranges of

\begin{equation}
\label{eq:sup:KP:GammaValues}
\gamma_1\in
\left[
	\frac{\pi}{2},\pi
\right],
\quad
\gamma_2 \in
\left[
	\frac{3\pi}{2},2\pi
\right].
\end{equation}Applying the same method as for $\mathcal{Q}_x$, the integrals evaluate to 
\begin{equation}
\mathcal{W}_x=
-e^{i\frac{3\pi}{4}}
e^{i\frac{(V_x-c_x)^2}{4\xi B}t}
\sqrt{\frac{\pi}{\xi B t}}.
\end{equation}

	\subsubsection{Evaluation of $\mathcal{W}_z$ sub-integral}
		The structure of the integral $\mathcal{W}_z$ is identical to that of $\mathcal{W}_x$ with $z$ instead of $x$. Therefore,

\begin{equation}
\mathcal{W}_z=
-e^{i\frac{3\pi}{4}}
e^{i\frac{(V_z-c_z)^2}{4\xi B}t}
\sqrt{\frac{\pi}{\xi B t}}
\end{equation}

\section{Supplemental Material: 2D-CMH: Analysis}
	\label{sup:NP:CMH}
	\subsection{The effect of initial conditions on the stability of 2D-CMH solutions}
		\label{sup:NP:InitialConditions}

Following the same steps as in Section~\ref{sup:NP:FourierIntegral}, the following is obtained for the solution to Eq.~\ref{eq:NP:Operator} in the algebraic domain (the inverse Laplace transform is not necessary for this step),	
			
\begin{equation}
\hat{H}
=
\frac{(A+U_0) + sH_0 + \eta H_0}
{s^2+\eta s+\eta^2},
\quad
\eta = k_x^2+k_z^2.
\end{equation}
Note that the forcing amplitude $A$ and the initial height disturbance $U_0$ have the exact same effect.

	\subsection{Development of the Fourier Inversion Integral for 2D-CMH operator}
		\label{sup:NP:FourierIntegral}
Taking Fourier transforms in $x$ and $z$ are taken of the 2D-CMH operator --Eq.~(\ref{eq:NP:Operator})-- yields a second order ordinary differential equation of the Fourier transformed variable, $\hat{\hat{h}}_{xz}(t)$.

\begin{equation}
\label{eq:sup:NP:FourierODE}
\frac{d^2 \hat{\hat{h}}_{xz}}{d t^2}
+
\Big(
\eta+B
\Big)\frac{d \hat{\hat{h}}_{xz}}{d t}
\\+
\Big(
\eta^2+B^2
\Big)\hat{\hat{h}}_{xz}
=A\delta(t).
\end{equation}	

\begin{equation}
\hat{\hat{h}}_{xz}(0)=0,
\quad
\frac{d\hat{\hat{h}}_{xz}}{dt}(0)=0,
\quad
\eta = k_x^2+k_z^2.
\end{equation}
Note that, in this construction, $\eta$ is always real and non-negative.
%

\noindent
Eq.~(\ref{eq:sup:NP:FourierODE}) is evaluated through a Laplace transform to yield the Fourier inversion integrals:

	\begin{equation*}
h(x,z,t) = 
\frac{A}{2\pi^2\sqrt{3}}
\int\limits_{-\infty}^{\infty}
\int\limits_{-\infty}^{\infty}
\Big(
e^{-\frac{\eta}{2}t}
\Big)
\frac{\sin
\left(\frac{\eta}{2}\sqrt{3} t
\right)}
{\eta}
e^{ik_xV_xt}
e^{ik_zV_zt}
dk_zdk_x,
\end{equation*}

\begin{equation}
\label{eq:sup:NP:FourierInversionIntegralCartesian}
\eta = k_x^2+k_z^2,
\quad
V_x = \frac{x}{t},
\quad
V_z = \frac{z}{t}.
\end{equation}
	
	\subsection{Transformation of inversion integral into polar coordinates}
		\label{sup:NP:PolarCoordinates}
Eq.~(\ref{eq:sup:NP:FourierInversionIntegralCartesian})  is converted into polar coordinates using the following definitions:

\begin{equation}
\xi = \sqrt{k_x^2 + k_z^2},
\quad
k_x = \xi \cos(\theta),
\quad
k_z = \xi \sin(\theta),
\quad
dk_xdk_z = \xi d\xi d\theta.
\end{equation}
Letting $\mathcal{B} = 
\cos(\theta)V_x
+\sin(\theta)V_z$ gives us

\begin{equation}
\label{eq:sup:NP:FourierInversionIntegralPolar}
h(x,z,t) = 
\frac{A}{2\pi^2\sqrt{3}}
\int\limits_{0}^{\infty}
\frac{
\sin\left(
\frac{\sqrt{3}}{2}\xi^2t\right)
}{\xi}
e^{-\frac{\xi^2}{2}t}
\left(
\int\limits_{0}^{2\pi}
e^{i\mathcal{B}\xi 
t}d\theta
\right)
d\xi .
\end{equation}
Once the equation is in this form, the inner integral in $\theta$ can be evaluated followed by that of the outer integral in $\xi$.
		
		\subsubsection{Evaluation of $\theta$ integral for the special case of $\nicefrac{V_z}{V_x}=0$}
			\label{sup:NP:SpecialCaseTheta}
In order for the saddle points --Eq.~(\ref{eq:app:NP:ThetaSaddles})-- to lie on the ends of the integration contour, $\nicefrac{V_z}{V_x}=0$. This is the case when either $V_z$ is zero and $V_x$ is non-zero or when $V_x$ goes to infinity for finite $V_z$. Note here that, in this limit, the term $\hat{V}=\sqrt{V_x^2+V_z^2}$ is equivalent to $V_x$.

\begin{figure}[ht!]
\centering
\includegraphics[keepaspectratio,width=6in]{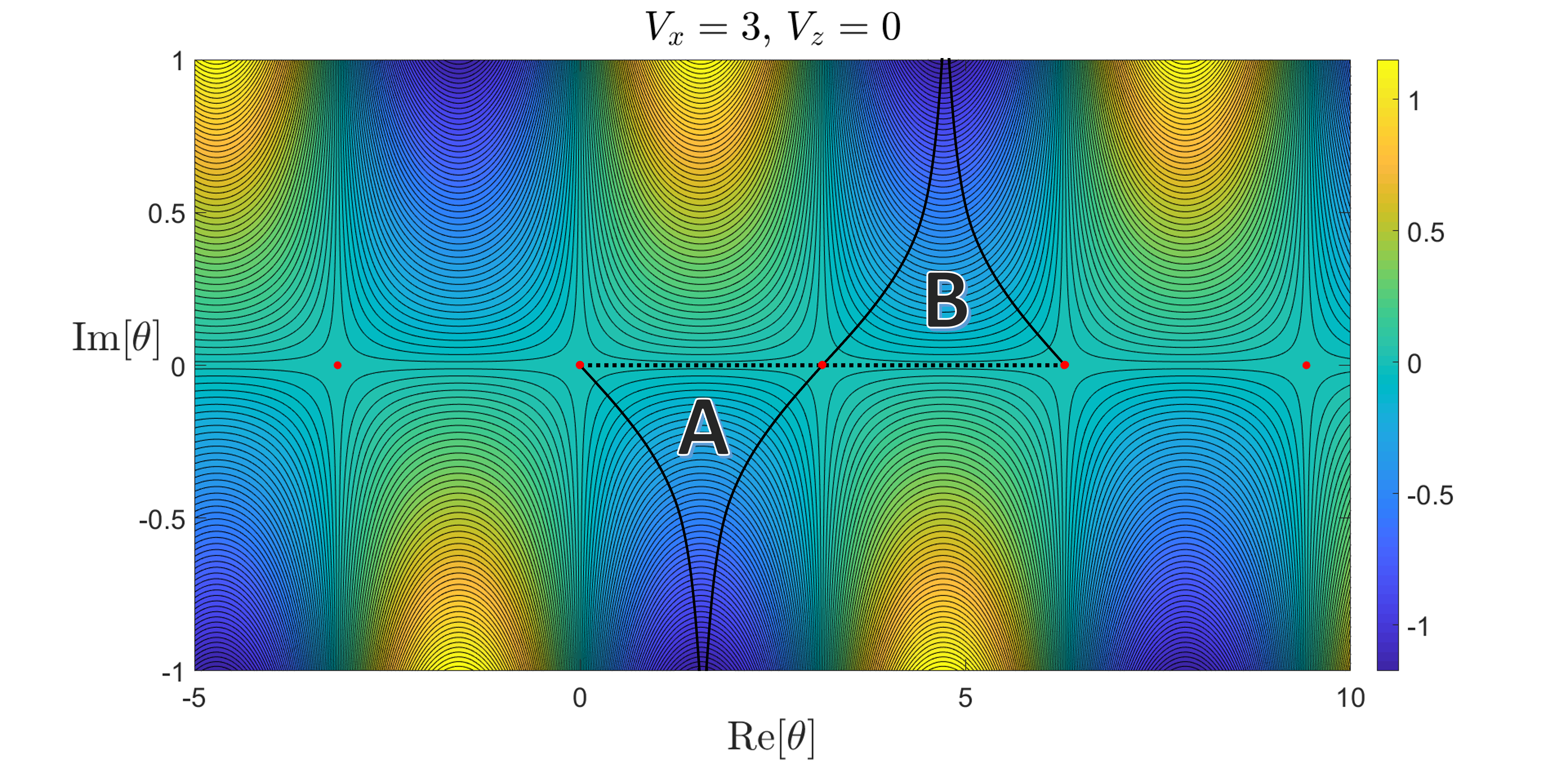}
\caption{Re$[i\Phi]$ vs Re$[\theta]$ and Im$[\theta]$. The integral of interest is marked with a dotted line running between $\theta = 0$ and $\theta = 2\pi$. Saddle points at $\theta_s= 0,\pi,2\pi$ are marked with (\textcolor{red}{$\bullet$}). Two closed contours are established which each contain a portion of the integral of interest. Each saddle point is at the maximum value of the contour, with each contour closing as Im$[\theta]\to\pm\infty$ which equates to Re$[i\Phi]\to -\infty$. Each contour will be generalized as two triangles in Figure~\ref{fig:sup:NP:TrianglesSpecialCase}.}
\label{fig:sup:NP:SaddleMapSpecialCase}
\end{figure}
\FloatBarrier

\noindent
For the following analysis, the contours are represented schematically as two triangles as shown below.	
	
\begin{figure}[ht!]
\centering
\includegraphics[keepaspectratio,width=4in]{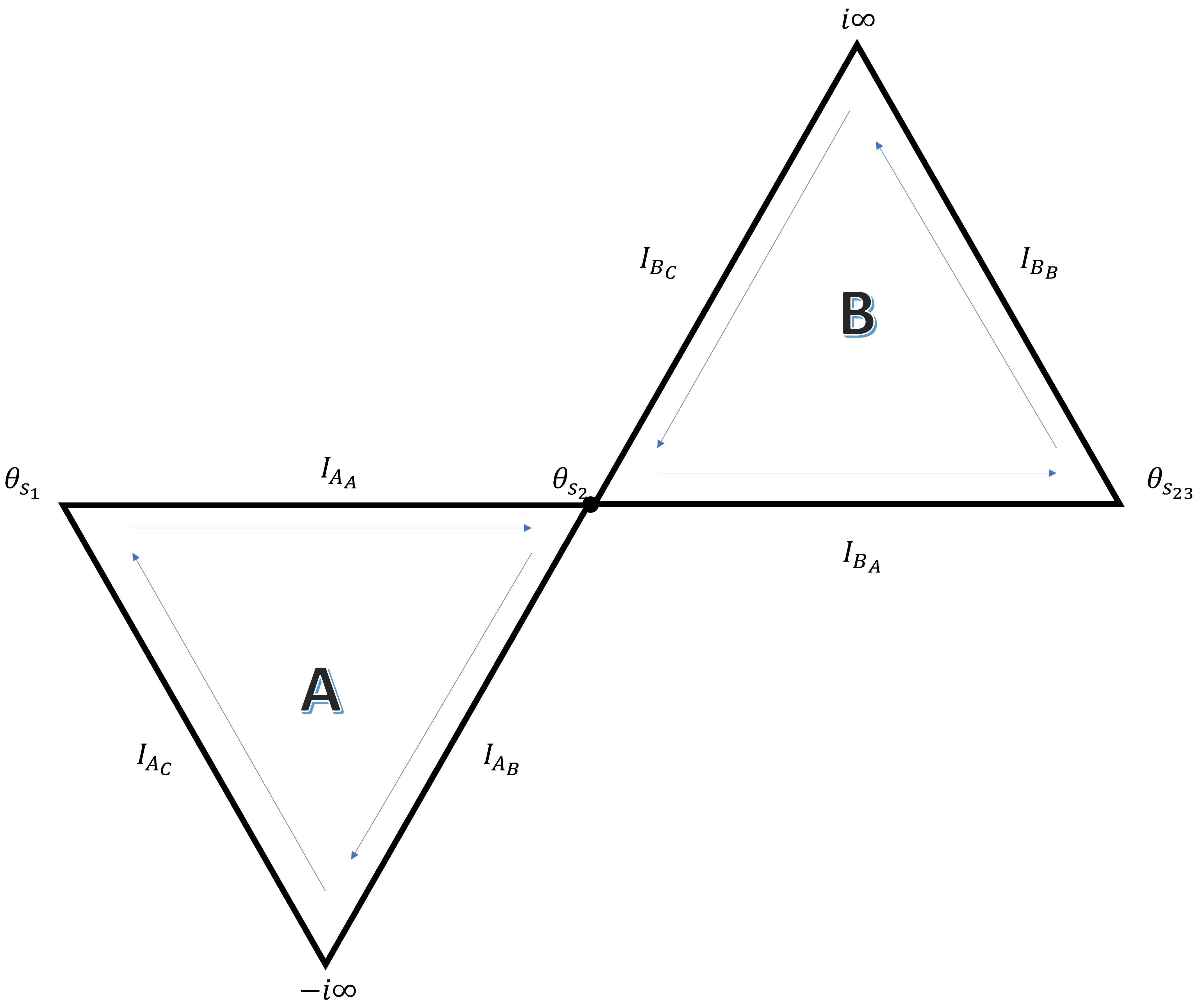}
\caption{simplified schematic of Figure~\ref{fig:sup:NP:SaddleMapSpecialCase}.
The integration contour for equation~(\ref{eq:app:NP:ThetaIntegral})  is given by the sub-contours $I_{A_A}$, $I_{B_A}$. The remaining sub-contours serve to form two closed contours, $A$ and $B$. The points $\theta_{S_1}$, $\theta_{S_2}$, and $\theta_{s_3}$ are three saddle points which lie on the integration contour. Note that the saddles lie at $\theta=(0,\pi,2\pi)$ when $\nicefrac{V_z}{V_x}=0$ as per equation~(\ref{eq:app:NP:ThetaSaddles}) .
}
\label{fig:sup:NP:TrianglesSpecialCase}
\end{figure}
\FloatBarrier

\noindent
Because there are no poles in or on the contours, each closed contour $A$ and $B$ can be written as

\begin{equation}
0=
\mathcal{I}_{A_A}
+\mathcal{I}_{A_B}
+\mathcal{I}_{A_C},
\end{equation}

\begin{equation}
0=
\mathcal{I}_{B_A}
+\mathcal{I}_{B_B}
+\mathcal{I}_{B_C}.
\end{equation}
Unlike the case in~\ref{app:NP:ThetaIntegral}, none of the contours can be neglected as time gets large. 
The integral $\mathcal{I}_\theta$ can be evaluated as

\begin{equation}
\label{eq:sup:NP:ThetaAsympFormGeneral}
\mathcal{I}_\theta\sim
-\mathcal{I}_{A_B}
-\mathcal{I}_{A_C}
-\mathcal{I}_{B_B}
-\mathcal{I}_{B_C}
\quad
\text{ as }
t\to\infty.
\end{equation}	
Eq.~(\ref{eq:sup:NP:ThetaAsympFormGeneral}) is evaluated through the same method as Eq.~(\ref{eq:app:NP:ThetaAsympFormGeneral}). Doing so yields

\begin{equation}
\mathcal{I}_\theta \sim
2
\Bigg(
\cos\left(V_x\xi t-\frac{\pi}{4}\right)
\Bigg)
	\sqrt{\frac{2\pi}
	{V_x\xi t}},
\end{equation}
which is equivalent to Eq.~(\ref{eq:NP:ThetaAsymptotic}).

\subsubsection{Conversion of $\xi$ integral into alternative integral $J$ for evaluation}
			\label{sup:NP:ConversionToJ}
			From Eq.~(\ref{eq:NP:XiIntegralAsymptotic}), we have
			
			\begin{equation}
			h\Big|_{\hat{V}\neq 0} \sim 
			\frac{A}{2\pi^2\sqrt{3}}
			\int\limits_0^{\infty}
			\frac{\sin
			\left(
			\frac{\sqrt{3}}{2}	
			\xi^2t
			\right)}{\xi}
			e^{-\frac{\xi^2}{2}t}
			\left(
			2\cos
			\left(
				\xi\hat{V}t
				-\frac{\pi}{4}
			\right)	\sqrt{\frac{2\pi}{\hat{V}\xi t}}
			\right)
			d\xi
			\qquad \text{ as } t\to\infty,
			\end{equation}
where $\hat{V}=\sqrt{V_x^2+V_z^2}$. 
Let $U = \xi^2t$, $dU = 2\xi t d\xi$, and thus	
			
			\begin{equation}
			h\Big|_{\hat{V}\neq 0} \sim 
			\left(
				\frac{(2\pi)^{\frac{1}{2}}}
				{\hat{V}^{\frac{1}{2}}t^{\frac{1}{4}}}
			\right)
			\int\limits_0^{\infty}
			\frac{\sin
			\left(
				\frac{\sqrt{3}}{2}
				U
			\right)
			}{
				U^{\frac{5}{4}}}
			e^{-\frac{U}{2}}
				\cos
				\left(
					\hat{V}(Ut)^\frac{1}{2}
					-\frac{\pi}{4}
				\right)
				dU
			\qquad \text{ as } t\to\infty.
			\end{equation}
				Rewriting the cosine as the real part of a complex exponential, we obtain

			\begin{equation}
			h\Big|_{\hat{V}\neq 0} \sim 
			\left(
				\frac{(2\pi)^{\frac{1}{2}}}
				{\hat{V}^{\frac{1}{2}}t^{\frac{1}{4}}}
			\right)
			\int\limits_0^{\infty}
			\frac{\sin
				\left(
					\frac{\sqrt{3}}{2}
					U
				\right)
				}{
				U^{\frac{5}{4}}}
			e^{-\frac{U}{2}}
			Re\Bigg[
				e^{
				\left(
					i\hat{V}(Ut)^\frac{1}{2}
					-\frac{i\pi}{4}
				\right)}
			\Bigg]
			dU
			\qquad \text{ as } t\to\infty,
			\end{equation}
				which can be rearranged as
	
			\begin{equation}
			h\Big|_{\hat{V}\neq 0} \sim 
			\left(
				\frac{(2\pi)^{\frac{1}{2}}}
				{\hat{V}^{\frac{1}{2}}t^{\frac{1}{4}}}
			\right)
			Re\Bigg[
			e^{
			\left(
				-i\frac{\pi}{4}
			\right)}
			\int\limits_0^{\infty}
			\frac{\sin
				\left(
					\frac{\sqrt{3}}{2}
					U
				\right)
				}{
				U^{\frac{5}{4}}}
			e^{-\frac{U}{2}}
			e^{
			\left(
				i\hat{V}(Ut)^\frac{1}{2}
			\right)}
			dU
			\Bigg]
			\qquad \text{ as } t\to\infty.
			\end{equation}
The result is Eqs.~(\ref{eq:NP:HInTermsOfJ}) and~(\ref{eq:NP:JIntegral}), and is transcribed below for completeness.

		\begin{equation}
			h\Big|_{\hat{V}\neq 0} \sim 
			\frac{A}{2\pi^2\sqrt{3}}
			\left(
				\frac{(2\pi)^{\frac{1}{2}}}
				{\hat{V}^{\frac{1}{2}}t^{\frac{1}{4}}}
			\right)
			Re\Bigg[
			e^{
			\left(
				-i\frac{\pi}{4}
			\right)}
			J(\hat{V},t)
			\Bigg]
			\qquad \text{ as } t\to\infty.
			\end{equation}
			
			\begin{equation}	\label{eq:sup:NP:JIntegralForm}
			J(\hat{V},t) = \int\limits_0^{\infty}
			\frac{\sin
				\left(
					\frac{\sqrt{3}}{2}
					U
				\right)
				}{
				U^{\frac{5}{4}}}
			e^{-\frac{U}{2}}
			e^{
			\left(
				i\hat{V}(Ut)^\frac{1}{2}
			\right)}
			dU
			\end{equation}

		\subsubsection{Evaluation of integral $J$}
			\label{sup:NP:JIntegral}
Starting from Eq.~(\ref{eq:sup:NP:JIntegralForm}), we complexfy the sine and separate it into two integrals, $J_1$ and $J_2$ according to Eq.~(\ref{eq:NP:JIntegrals}) but transcribed here for reference,

\begin{equation}
\label{eq:sup:NP:JIntegral1}
J_1 = \int\limits_0^{\infty}
\frac{\sin
	\left(
		\frac{\sqrt{3}}{2}
		W^2
	\right)
	}{
	W^{\frac{3}{2}}}
e^{-\frac{W^2}{2}}
	\cos(WS)
dW,
\end{equation}

\begin{equation}
\label{eq:sup:NP:JIntegral2}
J_2 = \int\limits_0^{\infty}
\frac{\sin
	\left(
		\frac{\sqrt{3}}{2}
		W^2
	\right)
	}{
	W^{\frac{3}{2}}}
e^{-\frac{W^2}{2}}
	\sin(WS)
dW,
\end{equation}
where $W = \sqrt{U}$ and $S=\sqrt{\hat{V}^2t}$.
In order to evalute~(\ref{eq:sup:NP:JIntegral1}), it is first differentiated with respect to $S$, complexified, and broken into two integrals

\begin{equation}
\frac{dJ_1}{dS} =
-\frac{(\mathcal{I}_1-\mathcal{I}_2)}{2i},
\end{equation}

\begin{equation}
\mathcal{I}_1 = 
\int\limits_0^{\infty}
\frac{e^{
	i\left(
		\frac{\sqrt{3}}{2}
		W^2
	\right)}
	}{
	W^{\frac{1}{2}}}
e^{-\frac{W^2}{2}}
	\sin(WS)
dW,
\end{equation}

\begin{equation}
\mathcal{I}_2 = 
\int\limits_0^{\infty}
\frac{e^{
	-i\left(
		\frac{\sqrt{3}}{2}
		W^2
	\right)}
	}{
	W^{\frac{1}{2}}}
e^{-\frac{W^2}{2}}
	\sin(WS)
dW.
\end{equation}
From WolframAlpha~\cite{WolframJ1Integral}

\begin{equation}
\mathcal{I}_1 = 
\frac{\pi S^\frac{1}{2} e^{-\frac{S^2}{8A_1}}
I_{\frac{1}{4}}
\left(
\frac{S^2}{8A_1}
\right)
}{
2\sqrt{2A_1}
},
\end{equation}

\begin{equation}
\mathcal{I}_2 = 
\frac{\pi S^\frac{1}{2} e^{-\frac{S^2}{8A_2}}
I_{\frac{1}{4}}
\left(
\frac{S^2}{8A_2}
\right)
}{
2\sqrt{2A_2}
},
\end{equation}
where

\begin{equation}
A_1 = \frac{1}{2}-i\frac{\sqrt{3}}{2},
\qquad
A_2 = \frac{1}{2}+i\frac{\sqrt{3}}{2}.
\end{equation}
Therefore,

\begin{equation}
\label{eq:sup:NP:J1Derivative}
\frac{dJ_1}{dS} =
-\frac{\pi S^\frac{1}{2}}{4i\sqrt{2}}
\left(
\frac{e^{-\frac{S^2}{8A_1}}
I_{\frac{1}{4}}
\left(
\frac{S^2}{8A_1}
\right)
}{\sqrt{A_1}}
-
\frac{e^{-\frac{S^2}{8A_2}}
I_{\frac{1}{4}}
\left(
\frac{S^2}{8A_2}
\right)
}{\sqrt{A_2}}
\right).
\end{equation}
The solution $J_1$ can be extracted from Eq.~(\ref{eq:sup:NP:J1Derivative}) by integration as

\begin{equation}
\label{eq:sup:NP:J1}
J_1=
\int\limits_{\infty}^{S}
\frac{dJ_1}{dS}\Big|_{S=\Omega}d\Omega.
\end{equation}
Note that, in order to evaluate this integral, the modified Bessel functions of the first kind are replaced with their asymptotic expansions for large arguments given by Eq.~(\ref{eq:NP:AllBesselInfo}). The resulting integrand is then integrated by parts to yield the asymptotic behaviour of $J_1$ at large $S$.
In order to evaluate~(\ref{eq:sup:NP:JIntegral2}), it is compexified and broken into two integrals

\begin{equation}
\label{eq:sup:NP:J2Form}
J_2 = \frac{J_{2_1}-J_{2_2}}{2i}
\end{equation}

\begin{equation}
\label{eq:sup:NP:J21Int}
J_{2_1} = \int\limits_0^{\infty}
\frac{e^{
	\left(
		i\frac{\sqrt{3}}{2}
		W^2
	\right)}
	}{
	W^{\frac{3}{2}}}
e^{-\frac{W^2}{2}}
	\sin(WS)
dW,
\end{equation}

\begin{equation}
\label{eq:sup:NP:J22Int}
J_{2_2} = 
\int\limits_0^{\infty}
\frac{e^{-
	\left(
		i\frac{\sqrt{3}}{2}
		W^2
	\right)
	}}{
	W^{\frac{3}{2}}}
e^{-\frac{W^2}{2}}
	\sin(WS)
dW.
\end{equation}
From WolframAlpha~\cite{WolframJ2Integral},

\begin{equation}
\int\limits_{0}^{\infty}
\xi^{-\frac{3}{2}}e^{-A\xi^2}
\sin( B\xi)d\xi
=
\frac{\pi B^{\frac{3}{2}} 
e^{-\frac{ B^2}{8A}}
\left(
I_{-\frac{1}{4}}\left(\frac{ B^2}{8A}\right)
+
I_{\frac{3}{4}}\left(\frac{ B^2}{8A}\right)
\right)}
{2\sqrt{2A}}.
\end{equation}
As before, we write

\begin{equation}
A_1 = \frac{1}{2}-i\frac{\sqrt{3}}{2},
\qquad
A_2 = \frac{1}{2}+i\frac{\sqrt{3}}{2}.
\end{equation}
The solutions to Eqs.~(\ref{eq:sup:NP:J21Int}) and~(\ref{eq:sup:NP:J22Int}) are

\begin{equation}
\label{eq:sup:NP:J21Sol}
J_{2_1} = 
\frac{\pi S^{\frac{3}{2}} 
e^{-\frac{ S^2}{8A_1}}
\left(
I_{-\frac{1}{4}}\left(\frac{ S^2}{8A_1}\right)
+
I_{\frac{3}{4}}\left(\frac{ S^2}{8A_1}\right)
\right)}
{2\sqrt{2A_1}},
\end{equation}

\begin{equation}
\label{eq:sup:NP:J22Sol}
J_{2_2} = 
\frac{\pi S^{\frac{3}{2}} 
e^{-\frac{ S^2}{8A_2}}
\left(
I_{-\frac{1}{4}}\left(\frac{ S^2}{8A_2}\right)
+
I_{\frac{3}{4}}\left(\frac{ S^2}{8A_2}\right)
\right)}
{2\sqrt{2A_2}}.
\end{equation}
Eqs.~(\ref{eq:sup:NP:J21Sol}) and~(\ref{eq:sup:NP:J22Sol}) are combined in Eq.~(\ref{eq:sup:NP:J2Form}) to yield

\begin{equation}
\label{eq:sup:NP:J2}
J_2 = 
\frac{\pi S^{\frac{3}{2}} }
	{4i\sqrt{2}}
\left(
	\frac{
	e^{-\frac{ S^2}{8A_1}}
	\left(
	I_{-\frac{1}{4}}\left(\frac{ S^2}{8A_1}\right)
	+
	I_{\frac{3}{4}}\left(\frac{ S^2}{8A_1}\right)
	\right)}
	{\sqrt{A_1}}
-
	\frac{
	e^{-\frac{ S^2}{8A_2}}
	\left(
	I_{-\frac{1}{4}}\left(\frac{ S^2}{8A_2}\right)
	+
	I_{\frac{3}{4}}\left(\frac{ S^2}{8A_2}\right)
	\right)}
	{\sqrt{A_2}}
\right)
\end{equation}

		\subsubsection{Cancellation of leading order terms of Bessel expansion in evaluation of $J$ integral}
		\label{sup:NP:OldLeadingOrder}
Equations~(\ref{eq:sup:NP:J1}) and~(\ref{eq:sup:NP:J2}) are evaluated using just the leading order terms of the Bessel expansion --Eq.~(\ref{eq:NP:AllBesselInfo})-- leading to the following result

\begin{equation}
J \sim \frac{2}{\sqrt{2\pi}} 
\Bigg(
	C_1	S^{-\frac{3}{2}}
\\
	+C_2 S^{-\frac{7}{2}}	
\\
	+C_3 S^{-\frac{15}{2}}
\Bigg)
\text{ as }
S\to\infty.
\end{equation}
The solution $J$ is written in terms of the aggregate constants

\begin{equation}
C_1 = 
	\frac{\sqrt{2}}{3}b\mathcal{F}_1 
	+ib(\mathcal{F}_1+\mathcal{G}_1),
\end{equation}

\begin{equation}
C_2 = 
	-\frac{2\sqrt{2}}{7}b\mathcal{F}_2 
	-2iab(\mathcal{F}_2+\mathcal{G}_2).
\end{equation}

\begin{equation}
C_3 = 
	-\frac{4\sqrt{2}}{15}(a^3b - ab^3)\mathcal{F}_4 
	-i(4a^3b - 4ab^3)(\mathcal{F}_4+\mathcal{G}_4),
\end{equation}

\begin{equation}
a=\frac{1}{2},
\quad
b=\frac{\sqrt{3}}{2}
\end{equation}
where the coefficients $\mathcal{F}_n$ and $\mathcal{G}_n$ are defined recursively as:
\begin{equation}
\mathcal{F}_1=
	(\frac{1}{4}-1),
\quad
\mathcal{F}_2=
	\mathcal{F}_1 \frac{(\frac{1}{4}-9)}{2},
\quad
\mathcal{F}_3=
	\mathcal{F}_2 \frac{(\frac{1}{4}-25)}{3},
\quad
\mathcal{F}_4=
	\mathcal{F}_3 \frac{(\frac{1}{4}-49)}{4},
\end{equation}

\begin{equation}
\mathcal{G}_1=
	(\frac{9}{4}-1)
\quad
\mathcal{G}_2=
	\mathcal{G}_1 \frac{(\frac{9}{4}-9)}{2},
\quad
\mathcal{G}_3=
	\mathcal{G}_2 \frac{(\frac{9}{4}-25)}{3},
\quad
\mathcal{G}_4=
	\mathcal{G}_3 \frac{(\frac{9}{4}-49)}{4}.
\end{equation}
As a note, the terms which contained just $\mathcal{F}_3$ and $\mathcal{G}_3$ cancel out exactly.
As stated in Eq.~\ref{eq:NP:CTermsZero},

\begin{equation}
Re\left[e^{-i\frac{\pi}{4}}C_1\right]=
Re\left[e^{-i\frac{\pi}{4}}C_2\right]=
Re\left[e^{-i\frac{\pi}{4}}C_3\right]=0.
\end{equation}
Therefore, we can conclude that $Re[	e^{-i\frac{\pi}{4}}J]=0$ if only the leading order terms are considered.	
		\subsubsection{Relevant terms of Bessel expansion to evaluate of $J$ integral}
			\label{sup:NP:NewLeadingOrder}
			
Equations~(\ref{eq:sup:NP:J1}) and~(\ref{eq:sup:NP:J2}) are evaluated using just the sub-dominant terms of the Bessel expansion --Eq.~(\ref{eq:NP:AllBesselInfo})--, and substituted into Eq.~(\ref{eq:NP:JIntegrals}) to yield

\begin{multline}
J\sim
\\  \sqrt{2\pi}e^{i\frac{\pi}{4}}
\Bigg(
	\sin\left(\frac{\sqrt{3}}{8} S^2\right)
	+\sqrt{3}\cos\left(\frac{\sqrt{3}}{8} S^2\right)
	-i\cos\left(\frac{\sqrt{3}}{8} S^2\right)
	+i\sqrt{3}\sin\left(\frac{\sqrt{3}}{8} S^2\right)
\Bigg)
S^{-\frac{3}{2}}
e^{-\frac{S^2}{8}}
\\+
	O\left(
		S^{-\frac{5}{2}}
		e^{-\frac{S^2}{8}}
	\right)\text{ as }
S\to\infty,
\end{multline}
where $S = \sqrt{\hat{V}^2t}$ and $\hat{V}=\sqrt{V_x^2+V_z^2}$

\end{document}